\documentclass[%
reprint,
superscriptaddress,
longbibliography,
amsmath,amssymb,
aps,
prl,
floatfix,
]{revtex4-2}

\usepackage{amsmath}
\usepackage{amssymb}
\usepackage{mathrsfs}
\usepackage{graphicx}
\usepackage{dcolumn}
\usepackage{xcolor}

\usepackage[colorlinks=true, linkcolor=blue, citecolor=blue, urlcolor=blue]{hyperref}
\newcommand{\pnl}[3][]{%
  \sbox0{\includegraphics[#1]{#3}}%
  \sbox2{\usebox0%
    \llap{\raisebox{\dimexpr\ht0-1.8ex\relax}{\makebox[\wd0][l]{\sffamily\small(#2)}}}}%
  \leavevmode\raisebox{-\height}{\usebox2}%
}

\begin{document}
\title{
Haldane--Holstein model at fractional filling: Route to bosonic fractional Chern insulator and quantum anomalous Hall crystal
}
\author{Zezhu Wei}
\affiliation{Department of Materials Science and Engineering, University of Washington, Seattle, Washington 98195, USA}

\author{Ang-Kun Wu}
\affiliation{Department of Physics and Astronomy, University of Tennessee, Knoxville, Tennessee 37996, USA}

\author{Di Xiao}
\affiliation{Department of Materials Science and Engineering, University of Washington, Seattle, Washington 98195, USA}
\affiliation{Department of Physics, University of Washington, Seattle, Washington 98195, USA}

\author{Shi-Zeng Lin}
\affiliation{Theoretical Division, T-4 and CNLS, Los Alamos National Laboratory, Los Alamos, New Mexico 87545, USA}
\affiliation{Center for Integrated Nanotechnologies (CINT), Los Alamos National Laboratory, Los Alamos, New Mexico 87545, USA}
\date{\today}

\begin{abstract}
Electron-phonon coupling is generally expected to suppress band topology, driving a topological band insulator into a trivial phase. Here we show that, using the prototypical Haldane--Holstein model at partial filling of a topological band, strong electron-phonon coupling can instead stabilize topological phases of matter. The electron-phonon coupling plays a double role: it generates the longer-range interactions that correlate the carriers and, in the attractive channel, also binds them into bosonic pairs. For spinful electrons, it pairs opposite spins into charge-$2e$ bipolarons that form a bosonic fractional Chern insulator (FCI) at filling $\nu=1/2$ with charge-$e$ semionic excitations. Tuning the band topology and the coupling strength maps out a rich phase diagram containing this bosonic FCI, superconductors condensed at the $M$ and $K$ points, and several charge-ordered solids.
For spin-polarized electrons, strong coupling instead stabilizes a $C=2$ quantum anomalous Hall crystal (QAHC) at $\nu=1/3$. Electron-phonon coupling thus emerges as a route to, rather than an obstruction against, topology in partially filled Chern bands.
\end{abstract}

\maketitle

{\it Introduction.---}
Topological bands with a nonzero Chern number have produced a stream of exciting findings in recent years, most prominently the observation of the fractional Chern insulator (FCI) in twisted MoTe$_2$~\cite{Cai2023:Signatures,Zeng2023:Thermodynamic,Park2023:Observation,Xu2023:Observation,Ji2024:Local,Redekop2024:Direct,Xu2025:Signatures} and rhombohedral multilayer graphene~\cite{Lu2024:Fractional,Lu2025:Extended,Xie2025:Tunable,Waters2025:Chern,Li2026:Fractional}. The FCI arises from the interplay between band topology and many-body interactions in a topological flat band. While this interplay has been extensively investigated in toy models~\cite{Tang2011:High,Sun2011:Nearly,Neupert2011:Fractional,Sheng2011:Fractional,Regnault2011:Fractional} and, more recently, in realistic materials~\cite{Xiao2011:Interface,Wu2019:Topological, Li2021:Spontaneous, Reddy2023:Fractional, Wang2024:Fractional, Ledwith2020:Fractional, Repellin2020:Chern, Liu2021:Gate}, the interactions responsible for FCIs have almost exclusively been attributed to electron-electron correlations. In contrast, the role of electron-phonon coupling remains far less explored, despite the ubiquity of lattice vibrations in quantum materials and their central role in phenomena ranging from superconductivity and charge-density waves to polaron formation. Understanding how electron-phonon coupling competes or cooperates with topology is therefore an important open question.
Earlier studies focused on the topological phase transition induced by Holstein-type electron-phonon coupling in topological lattice models at integer fillings, including those for Chern insulators~\cite{Cangemi2019:Topological,Islam2024:Electron,SousaJunior2026:Real} and
topological insulators~\cite{Cangemi2019:TopologicalPhase,Lu2023:Topological,Bhattacharyya2024:Holstein}.
These studies have shown that when one or more Chern bands are completely filled, sufficiently strong Holstein coupling destroys the Chern topology, driving a topological-to-trivial transition. Intuitively, the phonon-induced localization of electronic wave functions competes with the delocalized Bloch states required to support a nontrivial Chern band.

In this Letter we show that, in a partially filled topological band, strong electron-phonon coupling can induce rather than destroy topological phases. We work in the strong-coupling limit of the Haldane--Holstein model, the minimal and canonical setting for the interplay between topology and correlation~\cite{Cangemi2019:Topological,Islam2024:Electron,SousaJunior2026:Real}. Integrating out the phonons, we find that the coupling simultaneously (i) suppresses the electrons' kinetic energy, (ii) generates longer-range density-density interactions absent in the bare model, and (iii) in the attractive channel, binds the carriers into bosonic pairs.
For spinful electrons, these pairs are charge-$2e$ bipolarons, hard-core bosons that form a bosonic FCI at $\nu=1/2$. The bosonic FCI belongs to the family known from extended-Haldane hard-core-boson models~\cite{Wang2011:Fractional,Lu2025:Continuous,Lu2026:Vestigial}, but here the bosons and the interactions that fractionalize them are generated dynamically rather than imposed by hand. This gives an unusual hierarchy of binding and fractionalization: electron-phonon coupling first binds electrons into charge-$2e$ bipolarons, which subsequently fractionalize in the FCI into charge-$e$ semions, excitations with the charge of an electron but fractional quantum statistics~\cite{Halperin1984:Statistics,Arovas1984:Fractional,Feldman2021:Fractional}. 
Notably, no flat electronic band is required: the flat Chern band hosting the FCI is emergent for bipolarons, while the electronic band itself stays strongly dispersive.
Tuning the band topology and the coupling strength maps out a rich phase diagram containing the bipolaron FCI, superconductors at the $M$ and $K$ points, and several bipolaron solids.
For spin-polarized electrons, the same model instead yields a $C=2$ quantum anomalous Hall crystal (QAHC) at filling $\nu=1/3$, demonstrating that strong electron-phonon coupling also provides an alternative microscopic mechanism for realizing a QAHC~\cite{Zhou2024:Fractional, Dong2024:Anomalous, Kwan2025:Moire, Tan2024:Parent, Sheng2024:Quantum, Dong2024:Stability, Zeng2024:Sublattice, Goncalves2025:Doping,PereaCausin2025:Quantum}.

\begin{figure}[t]
    \centering
    \pnl{a}{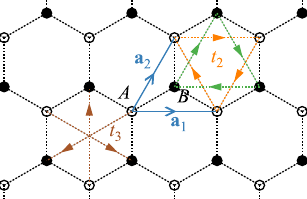}\\[2pt]
    \pnl{b}{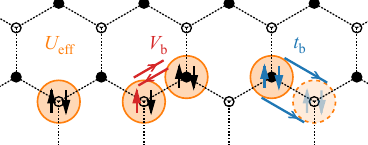}\\[2pt]
    \pnl{c}{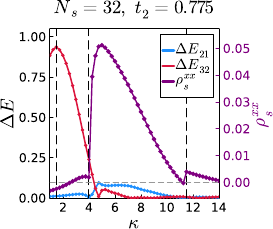}\hfill
    \pnl{d}{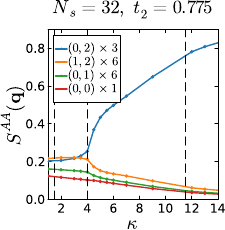}%
    \caption{(a) Schematic of the extended Haldane model. (b) Schematics of bipolaron formation, and of the interaction $V_b$ and hopping $t_b$ of bipolarons; the phonon cloud is represented by the shaded circle. (c,d) Exact-diagonalization (ED) results for the effective hard-core boson Hamiltonian at $\nu=1/2$ and maximal flatness ratio of the bipolaron band, on $N_s=32$ sites, with $\kappa$ setting the interaction-to-hopping ratio; the dashed lines mark $\kappa = 1.5$, $4$, and $11.5$. (c) Ground-state splitting $\Delta E_{21}=E_2-E_1$ and many-body gap $\Delta E_{32}=E_3-E_2$ (left axis), and superfluid stiffness $\rho_{s}^{xx}$ (right axis) against $\kappa$. (d) Intra-sublattice structure factor $S^{AA}(\mathbf{q})$ against $\kappa$.}
    \label{fig:fig1}
\end{figure}

{\it Haldane--Holstein model for spinful electrons.---}
We consider a generalized Haldane--Holstein model on the honeycomb lattice~\cite{Cangemi2019:Topological,SousaJunior2026:Real}, combining an extended Haldane model~\cite{Haldane1988:Model} with the Holstein--Hubbard model~\cite{Holstein1959:Studies},
\begin{align}\label{eq:spinful-orig}
    &\hat{H} =  - t_1 \sum_{\langle i,j \rangle, \sigma} \hat{c}_{i \sigma}^{\dagger} \hat{c}_{j \sigma} 
    - t_{2}\sum_{\langle\langle i,j\rangle\rangle, \sigma}e^{i \nu_{ij}\phi} \hat{c}_{i \sigma}^{\dagger} \hat{c}_{j \sigma} \nonumber\\
    &\quad- t_{3}\sum_{\langle\langle\langle i,j\rangle\rangle\rangle, \sigma}e^{i \mu_{ij} \phi_2} \hat{c}_{i \sigma}^{\dagger} \hat{c}_{j \sigma} + \frac{U_{\text{e-e}}}{2} \sum_{i} \hat{n}_{i}^{2} \nonumber\\
    &\quad+ \alpha \sum_{i} \hat{n}_{i} \hat{x}_{i} 
    + \sum_{i} \left(\frac{\hat{p}_{i}^{2}}{2M} + \frac{M \omega^2 \hat{x}_{i}^{2}}{2}\right),
\end{align}
where $\langle i,j \rangle$, $\langle \langle i,j \rangle \rangle$, $\langle \langle \langle i,j \rangle \rangle \rangle$ stand for pairs of nearest-neighbor (NN), second-nearest-neighbor (2NN), and third-nearest-neighbor (3NN) sites, $\sigma=\uparrow,\downarrow$ labels the electron spin, $\hat{n}_{j} = \sum_{\sigma} \hat{c}^\dagger_{j \sigma} \hat{c}_{j \sigma}$ is the total electron number operator, and $\hat{x}_i$ and $\hat{p}_i$ are conjugate position and momentum operators for phonons.
We set $t_1 = 1$ as the unit of energy and $\hbar = 1$.
The hopping phases $\phi,\ \phi_2\neq n \pi$ with an integer $n$ break time-reversal symmetry, with $\nu_{ij} = +1$ $(-1)$ for hopping $j\to i$ along (against) the arrows in Fig.~\ref{fig:fig1}(a); similarly, $\mu_{ij}=+1$ $(-1)$ for 3NN hopping from sublattice A to B (B to A).

The electron-phonon coupling term can be formally eliminated by the Lang--Firsov transformation~\cite{LangFirsov}, $\hat{U}=\prod_{i} \exp[i(\alpha/ M \omega^2) \hat{p}_{i} \hat{n}_{i}] $ [End Matter, Eq.~(\ref{eq:spinful-transformed})].
In exchange, two of the surviving terms are modified.
First, an on-site electron attraction $-U_{\text{ph}} \sum_{i} \hat{n}_i^2/2$ is added to the Hubbard repulsion, with $U_{\text{ph}} = \alpha^2/(M \omega^2)$, leaving a net on-site interaction $U_{\text{eff}} = U_{\text{e-e}} - U_{\text{ph}}$, whose sign decides the physics.
Second, every hopping is dressed by a phonon operator $\hat{S}_{ij}$, whose zero-phonon average $\langle 0 |\hat{S}_{ij}| 0 \rangle = e^{-X/2}$, the Franck--Condon factor~\cite{Carlson2008:Concepts}, suppresses the bare amplitude. 
Here $X = U_{\text{ph}}/\omega = \alpha^{2}/(M \omega^{3})$
is a dimensionless electron-phonon coupling parameter.
We work in the strong-coupling regime, where the dressed hoppings are weak and can be treated perturbatively~\cite{Han2020:Strong}.
This regime divides into two cases, $U_{\text{ph}} > U_{\text{e-e}}$ and $U_{\text{ph}} < U_{\text{e-e}}$; we treat the former first.

\begin{figure}[t]
    \centering
    \includegraphics{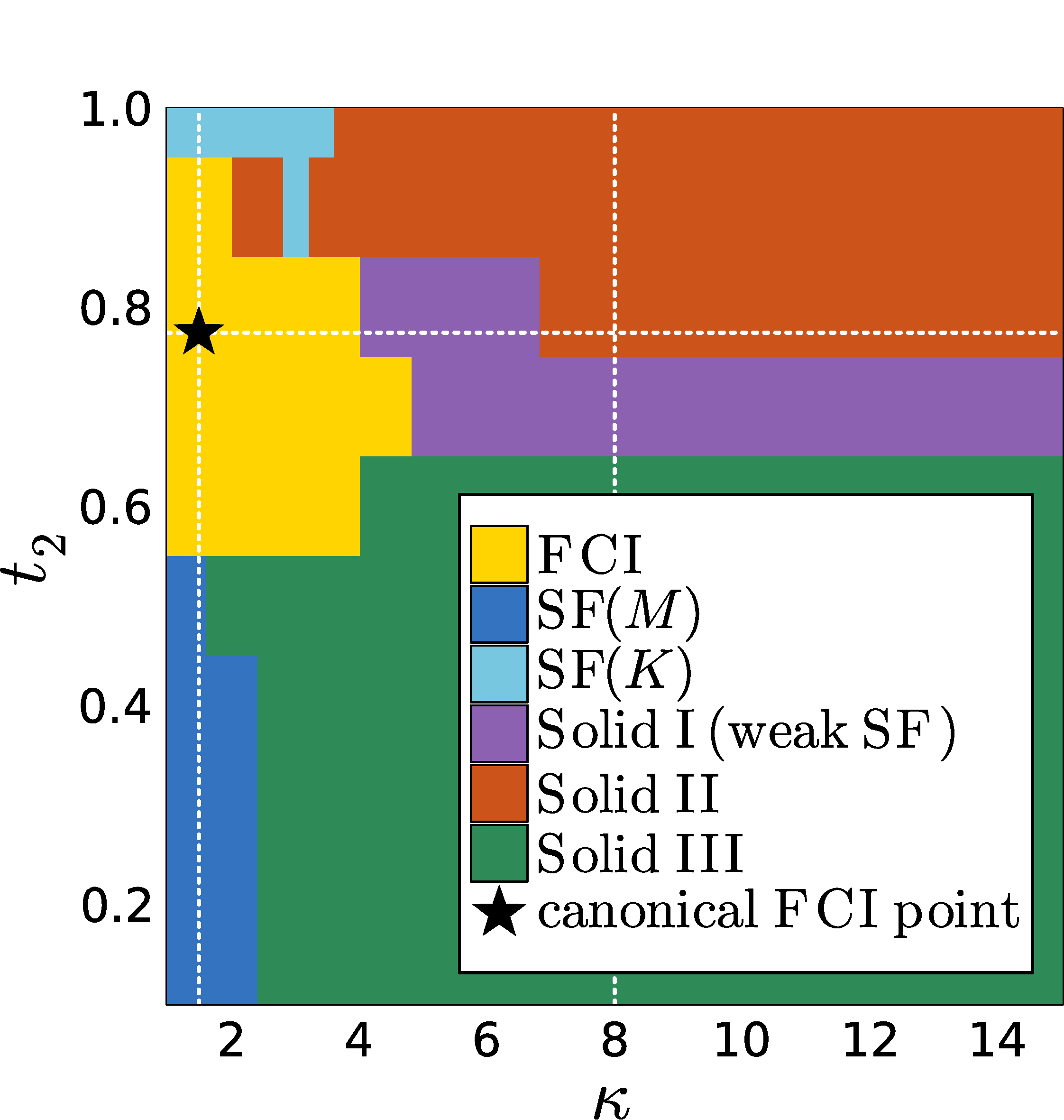}%
    \caption{Phase diagram of the effective hard-core boson Hamiltonian at $\nu=1/2$ on $N_s=24$ sites in the $(\kappa, t_2)$ plane. The dotted lines are the three cuts $\kappa=1.5$, $\kappa=8$ and $t_2=\sqrt{0.6}$ along which the phases were identified. The order of the phase transition cannot be resolved within the small-system ED calculations.}
    \label{fig:fig2}
\end{figure}

\begin{figure}[t]
    \centering
    \pnl[scale=0.97]{a}{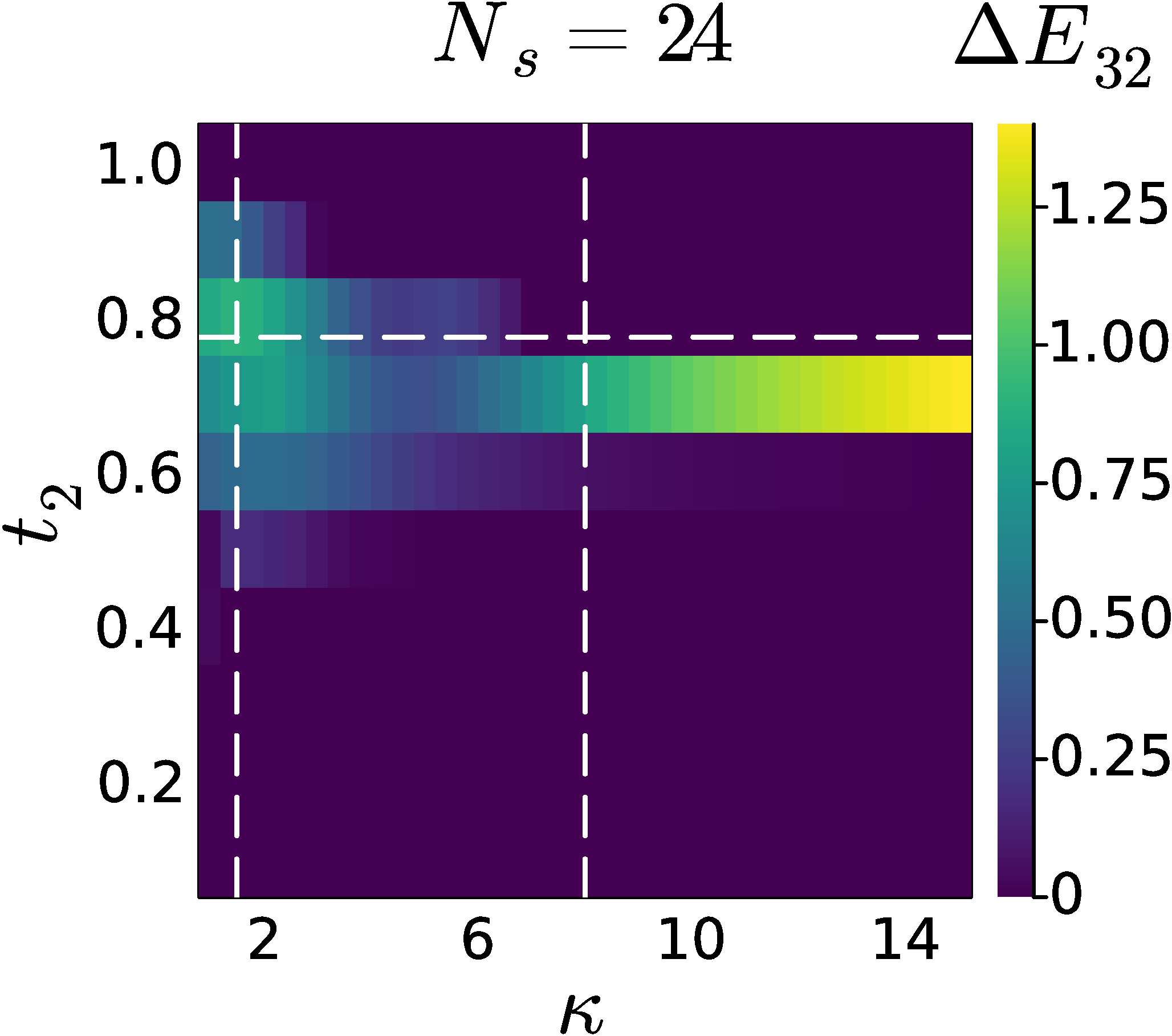}\hfill
    \pnl[scale=0.97]{b}{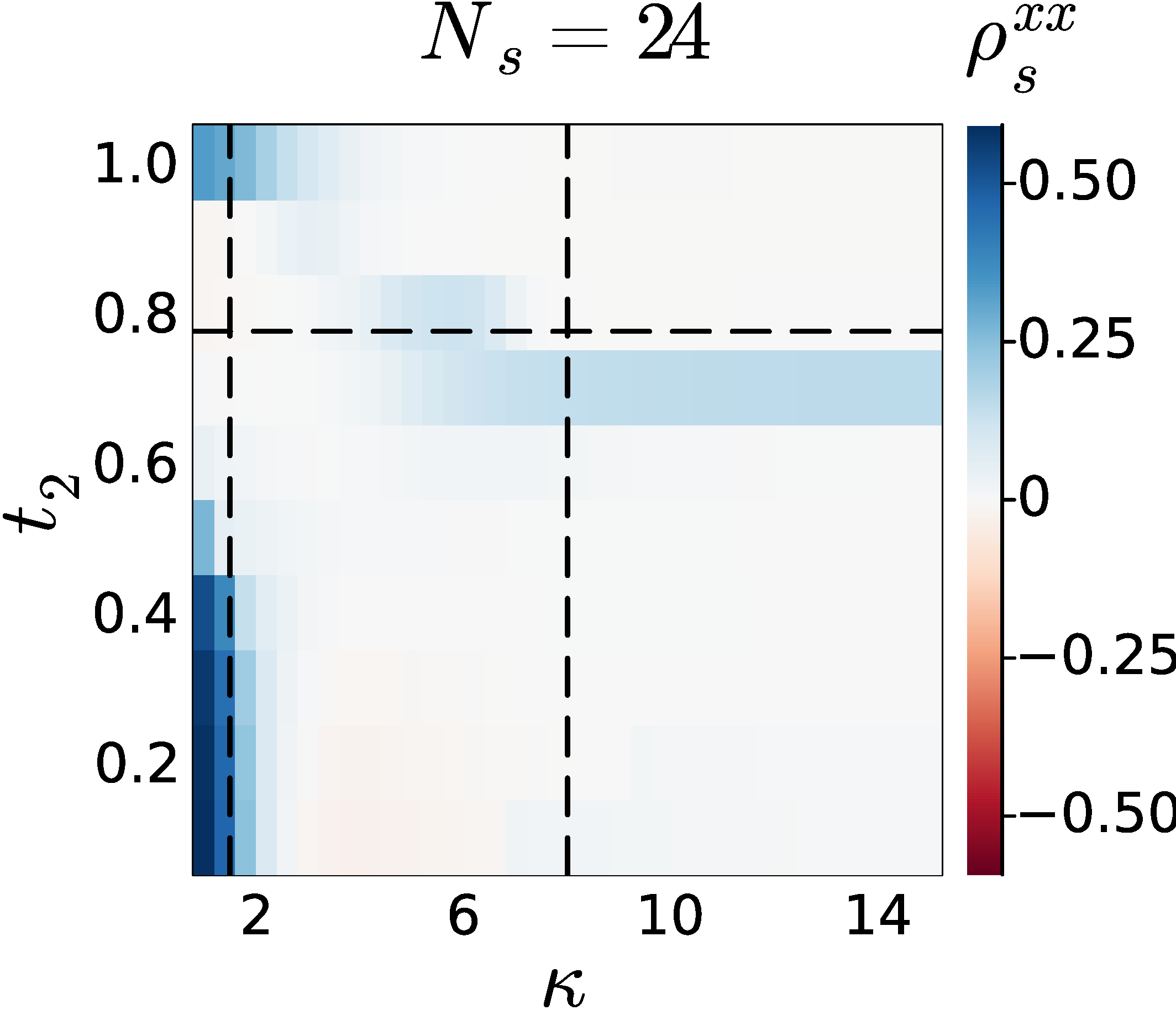}\\[2pt]
    \pnl{c}{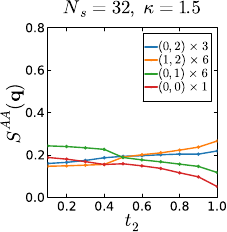}\hspace{6pt}%
    \pnl{d}{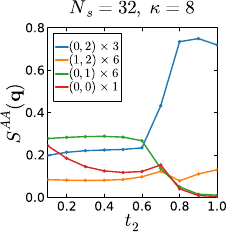}%
    \caption{ED results for the effective hard-core boson Hamiltonian at $\nu=1/2$, treating both $t_2$ and $\kappa$ as free parameters.
    (a,b) Many-body gap $\Delta E_{32} $ and superfluid stiffness $\rho_{s}^{xx}$ of the ground states on $N_s=24$ sites. The dashed lines are three cuts $\kappa = 1.5$, $\kappa = 8$, and $t_2 = \sqrt{0.6}$.
    (c,d) Intra-sublattice structure factor $S^{AA}(\mathbf{q})$ of the ground states on $N_s=32$ sites against $t_2$, for $\kappa=1.5$ and $\kappa=8$.}
    \label{fig:fig3}
\end{figure}

{\it Bipolarons at $U_\mathrm{eff}<0$.---}
When $U_{\text{ph}} > U_{\text{e-e}}$, the net on-site interaction is attractive despite the Hubbard repulsion: strong electron-phonon coupling binds electrons of opposite spins with a binding energy $|U_{\text{eff}}|$.
These on-site pairs, $\hat{b}_i = \hat{c}_{i \downarrow} \hat{c}_{i \uparrow}$, are dressed by phonons, forming bipolarons [Fig.~\ref{fig:fig1}(b)].
We carry out a degenerate perturbation expansion in the hopping terms to second order~\cite{Han2020:Strong}.
The ground-state manifold of the unperturbed Hamiltonian $\hat{H}_0 = ({U_{\text{eff}}}/{2}) \sum_{i} \hat{n}_{i}^{2}
+ \omega \sum_{i} \hat{a}_i^\dagger \hat{a}_i$, with $\hat{a}_i$ the phonon annihilation operator, consists of states in which each site is either empty or occupied by a bipolaron, with no phonon excitations.
Integrating out the phonons in the excited states leaves a hard-core boson model on the honeycomb lattice [End Matter, Eq.~(\ref{eq:eff_ham})], in which the bipolarons hop between NN, 2NN and 3NN sites with amplitudes $t_b$ and repel each other at the same three ranges with strengths $V_b$.
Both descend from the same virtual pair breaking [Fig.~\ref{fig:fig1}(b)]: one electron of a bipolaron hops to a neighboring site and either returns or is followed by its partner.
If the neighbor hosts another bipolaron, it fills both spin states and blocks the virtual hop, and this blocked kinetic energy is the repulsion $V_b \propto t_{1,2,3}^{2}F(X,Y)/|U_{\text{eff}}|$, absent in Eq.~(\ref{eq:spinful-orig}) and generated by the electron-phonon coupling.
If the partner follows instead, the pair moves with $t_b \propto t_{1,2,3}^{2}F(-X,Y)/|U_{\text{eff}}|$, but must drag its phonon cloud along, whose amplitudes partially cancel, so $F(-X,Y) < F(X,Y)$ and stronger coupling favors interaction over hopping.
Here $F(x,y) =  e^{-|x|} {}_1 F_{1}(y; y+1; x)$, with $_{1}F_{1}$ the confluent hypergeometric function, and $Y=|U_{\text{eff}}|/\omega = (U_{\text{ph}}-U_{\text{e-e}})/ \omega$.
Because every bipolaron amplitude is second order in the electron hopping, the bipolaron band is narrow, and the doubled phases $2\phi$ and $2\phi_{2}$ on the 2NN and 3NN hoppings give it the $C=1$ topology of the original spinful model.

{\it Bipolaron FCI and phase transitions.---}
To map out the phases of this effective bosonic model, we perform exact diagonalization (ED) of the effective Hamiltonian on a torus of $N_s = 2 N_{x}  N_{y}$ sites at boson filling $\nu = N_{b}/(N_{x} N_{y})=1/2$, where $N_{b}$ is the number of bipolaron bosons; the original electrons are at quarter-filling of the lattice sites per spin.
Momentum vectors $\mathbf{k} = ({k_x} /{N_x}) \mathbf{b}_1  + ({k_y}/{N_y}) \mathbf{b}_2$ are labeled by the integer quantum numbers $(k_x, k_y)$, where $\mathbf{b}_{1,2}$ are the reciprocal lattice vectors. See the End Matter for lattice and observable conventions.

We first choose parameters that maximize the flatness ratio of the lower band, that is, the ratio of the band gap to the bandwidth~\cite{Wang2011:Fractional}: $t_2/t_1=\sqrt{0.6}$, $t_3/t_1=\sqrt{0.58}$, $\phi=0.2\pi$, and $\phi_2=0.5\pi$.
These values flatten the bipolaron band rather than the electronic one, giving it a flatness ratio of $\approx 50$.
The only remaining tuning parameter is then $\kappa = F(X,Y)/{F}(-X,Y)$, which fixes the effective interaction-to-hopping ratio to $2\kappa$.
This ratio grows with the coupling strength $\alpha$ and with decreasing phonon frequency $\omega$; note $\kappa>1$ since $F(X,Y) > {F}(-X,Y)$.

Figures~\ref{fig:fig1}(c,d) track four quantities as $\kappa$ increases: the many-body gap $\Delta E_{32}$, the ground-state splitting $\Delta E_{21}$, the superfluid stiffness $\rho_s^{xx}$, and the intra-sublattice structure factor $S^{AA}(\mathbf{q})$. In Figs.~\ref{fig:fig1}(d) and \ref{fig:fig3}(c,d), symmetry-equivalent momenta $\mathbf{q}$, labeled by integers $(q_x, q_y)$, are merged and $\times N$ denotes their multiplicity.
As we increase the ratio $\kappa$, the system passes from an FCI phase, first to a Solid I phase with weak superfluidity, and finally to a Solid II phase. Similar interaction-driven transitions in hard-core bosons were reported in earlier works~\cite{Wang2011:Fractional,Luo2020:Quantum,Lu2026:Vestigial}, though with different interactions and tuning parameters.

Between $\kappa=1$ and $4$, the many-body spectrum exhibits two degenerate ground states, at momentum sectors $(k_x, k_y) = (0,0)$ and $(0,2)$ for $N_s = 24$ and both at $(0,0)$ for $N_s = 32$, in agreement with the $1/2$ bosonic FCI counting.
At $\kappa=1.5$, the spectral flow under $y$-direction flux insertion shows the two ground states exchanging for $N_s=24$ and avoiding crossing for $N_s=32$, consistent with both residing in the same momentum sector [End Matter, Figs.~\ref{fig:fig5}].
We also compute the total many-body Chern number $C_{\text{tot}}$ of the ground-state manifold under twisted boundary phases $\theta_x$ and $\theta_y$, from its Berry curvature on a $10 \times 10$ grid.
For both $N_s=24$ and $32$, $C_{\text{tot}}=1$, and hence each ground state has many-body Chern number $1/2$. Altogether, these establish the bipolaron FCI. The quantized Hall conductance is $\sigma_{xy}=\nu(2e)^2/h=2{e^2}/{h}$. Correspondingly, the elementary quasiparticles carry charge $e$, i.e., one-half of the charge of the constituent bipolarons.

Between $\kappa=4$ and $\kappa=11.5$, the many-body gap $\Delta E_{32}$ closes, the ground-state splitting $\Delta E_{21}$ grows, and a finite superfluid stiffness appears, signaling a phase transition at $\kappa=4$.
The transition is also accompanied by the onset of structure factor peaks at $\mathbf{b}_{2}/2$ and the other equivalent $M$ points [Fig.~\ref{fig:fig1}(d)], indicating a solid phase with weak superfluidity.
Such a phase may be a supersolid; however, two previous density matrix renormalization group (DMRG) studies disagree on whether a similar phase is a supersolid~\cite{Luo2020:Quantum,Lu2026:Vestigial}, so we call it Solid I.
When $\kappa>11.5$, the superfluid stiffness vanishes, the ground states are degenerate on the finite-size lattice, and the structure factor peaks at $\mathbf{b}_{2}/2$ continue to grow, signaling another transition into a Solid II phase.
However, no other peaks emerge across the Solid I--Solid II transition, in contrast to the similar sequence in Ref.~\cite{Lu2026:Vestigial}, where that transition shows growing peaks at $\pm (\mathbf{b}_1/4+ \mathbf{b}_2/2)$. This difference suggests the Solid II phases in the two scenarios have different origins; indeed, the NN interaction $V_1$ in Ref.~\cite{Lu2026:Vestigial} is held fixed at $4$, while all interactions grow with $\kappa$ here.

After mapping out the phases at maximal flatness ratio, we now relax that constraint by varying the time-reversal-symmetry-breaking 2NN hopping $t_2$ together with the ratio $\kappa$, again on $N_s=24$ and $32$ sites.
In Fig.~\ref{fig:fig2}, we show the resulting phase diagram in the $(\kappa, t_2)$ plane, and in Fig.~\ref{fig:fig3}, the supporting many-body gap, superfluid stiffness, and structure factor.
The phase diagram is now much richer; in addition to the FCI, Solid I, and Solid II of the $t_{2}=\sqrt{0.6}$ line, we further identify superfluids condensed at the $M$ and $K$ points [SF$(M)$ and SF$(K)$], and a Solid III phase with weaker charge order. Since the bipolarons carry charge $2e$, these superfluids are charge-$2e$ superconductors.
Along the line $\kappa = 1.5$, where the phonon coupling is relatively weak, we find an SF($M$)--FCI--SF($K$) sequence; a similar sequence was reported in Ref.~\cite{Lu2025:Continuous}. The bipolaron FCI is stabilized around $t_{2} = 0.8$, where Fig.~\ref{fig:fig3}(a) shows a finite gap above the doubly degenerate ground-state manifold.
For $t_2 < 0.6$ and $t_2 > 0.9$, the ground states are superfluids with finite superfluid stiffness in Fig.~\ref{fig:fig3}(b) and featureless structure factors in Fig.~\ref{fig:fig3}(c). The momentum occupation number $n(\mathbf{k})$ shows the two superfluids condensing at different momenta: at the $M$ points for $t_2 < 0.6$ and at the $K$ points for $t_2 > 0.9$ [End Matter, Fig.~\ref{fig:fig5}].
Along the line $\kappa=8$, for $t_{2} \approx 0.7$ and $t_2 > 0.8$, the ground states correspond to the previously identified Solid I and Solid II phases.
For $t_{2}<0.6$, the ground state is a third solid, Solid III, with vanishing superfluid stiffness and less prominent structure factor peaks at $\pm\mathbf{b}_{1,2}/4$ and $\pm(\mathbf{b}_1+\mathbf{b}_2)/4$ [Fig.~\ref{fig:fig3}(d)].

{\it Polarons in spin-polarized Haldane--Holstein model.---}
When both the on-site repulsion $U_\text{e-e}$ and the phonon-mediated attraction $U_{\text{ph}}$ are strong and $U_\text{eff}>0$,
the ground-state manifold of $\hat{H}_0$ consists of states without doubly occupied sites and without phonon excitations, describing polarons, i.e., electrons dressed by phonons.
In contrast to the $U_\text{eff}<0$ case, the first-order contribution does not vanish: it is simply the original Haldane model with every hopping suppressed by the Franck--Condon factor $e^{-X/2}$.
At second order the spinful problem generates Heisenberg spin couplings and spin-singlet-pair hopping, i.e., a $t$-$J$-$V$ model already studied in the adiabatic and intermediate regimes~\cite{Han2020:Strong,Huang2022:Pair}. 
Here we focus on the spin-polarized limit, where the spin sector is frozen and the interplay between topology and electron-phonon coupling is most transparent. No site can then be doubly occupied, so the on-site repulsion $U_{\text{e-e}}$ contributes only a constant at fixed filling, and we drop the $t_3$ term for simplicity. Beyond the $e^{-X/2}$ suppression, second-order perturbation theory on $\hat{H}_0 = \omega \sum_{i} \hat{a}_i^\dagger \hat{a}_i$ generates two further classes of terms [End Matter, Eq.~(\ref{eq:polaron-second})]: density-assisted hoppings, in which an electron moving between two sites picks up a factor $1-2\hat n_m$ from a middle bridging site $m$, and NN and 2NN density-density interactions $\propto t_{1,2}^{2}F'(X)/U_{\text{ph}}$.
Both are weighted by the phonon overlap sum $F'(x) =  x e^{-x} (\operatorname{Ei}(x)-\ln{x}-\gamma)$, with $\operatorname{Ei}(x)$ the exponential integral and $\gamma$ Euler's constant.

The electron-phonon coupling enters through the phonon frequency $\omega$ and the attraction $U_{\text{ph}}$, which we trade for the equivalent parameters $X$ and $U = U_{\text{ph}}/F'(X/2)$.
In the limit $X \to \infty$, since $F'(X) \to 1$, the density-density interactions dominate.
In the limit $U \to \infty$, both density-assisted hopping and density-density interactions vanish, and the model reduces to the noninteracting bare Haldane model, up to the overall factor $e^{-X/2}$.
In the intermediate regime, where both $U$ and $X$ are finite,
we use ED to compute the ground-state properties at the fermion fillings $\nu = N_{f}/(N_{x} N_{y})=1/3$ and $2/3$, where $N_{f}$ is the fermion number.
We fix the band topology using the parameters that maximize the flatness ratio of the bare Haldane band~\cite{Neupert2011:Fractional}, $ t_2/t_1=(\sqrt{{43}/{3}})/{12 }$ and $\phi=\pi+\arccos(3 \sqrt{{3}/{43}})$, for which that ratio is $\approx 6$.

\begin{figure}[t]
    \noindent
    \pnl{a}{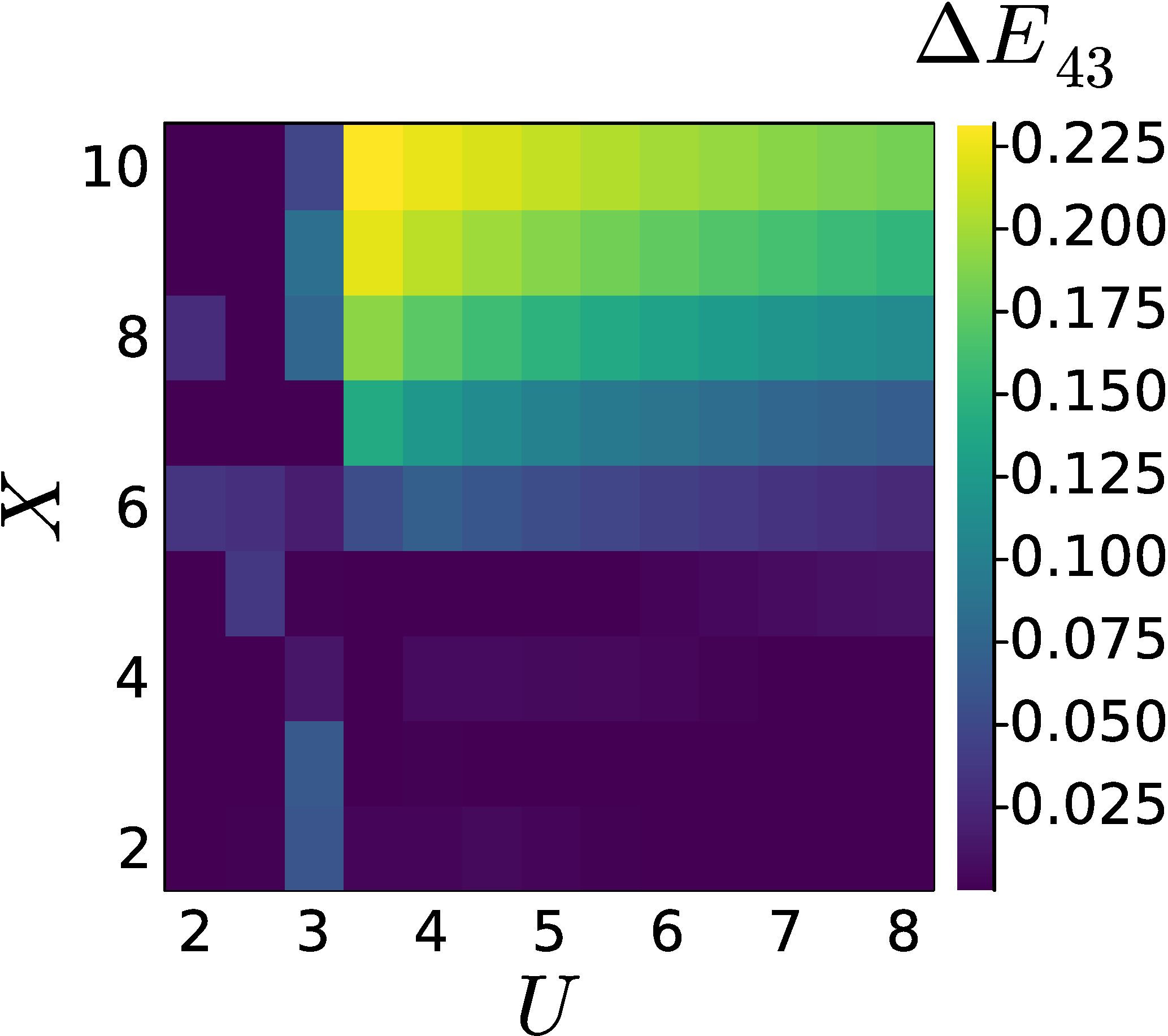}\hfill
    \pnl{b}{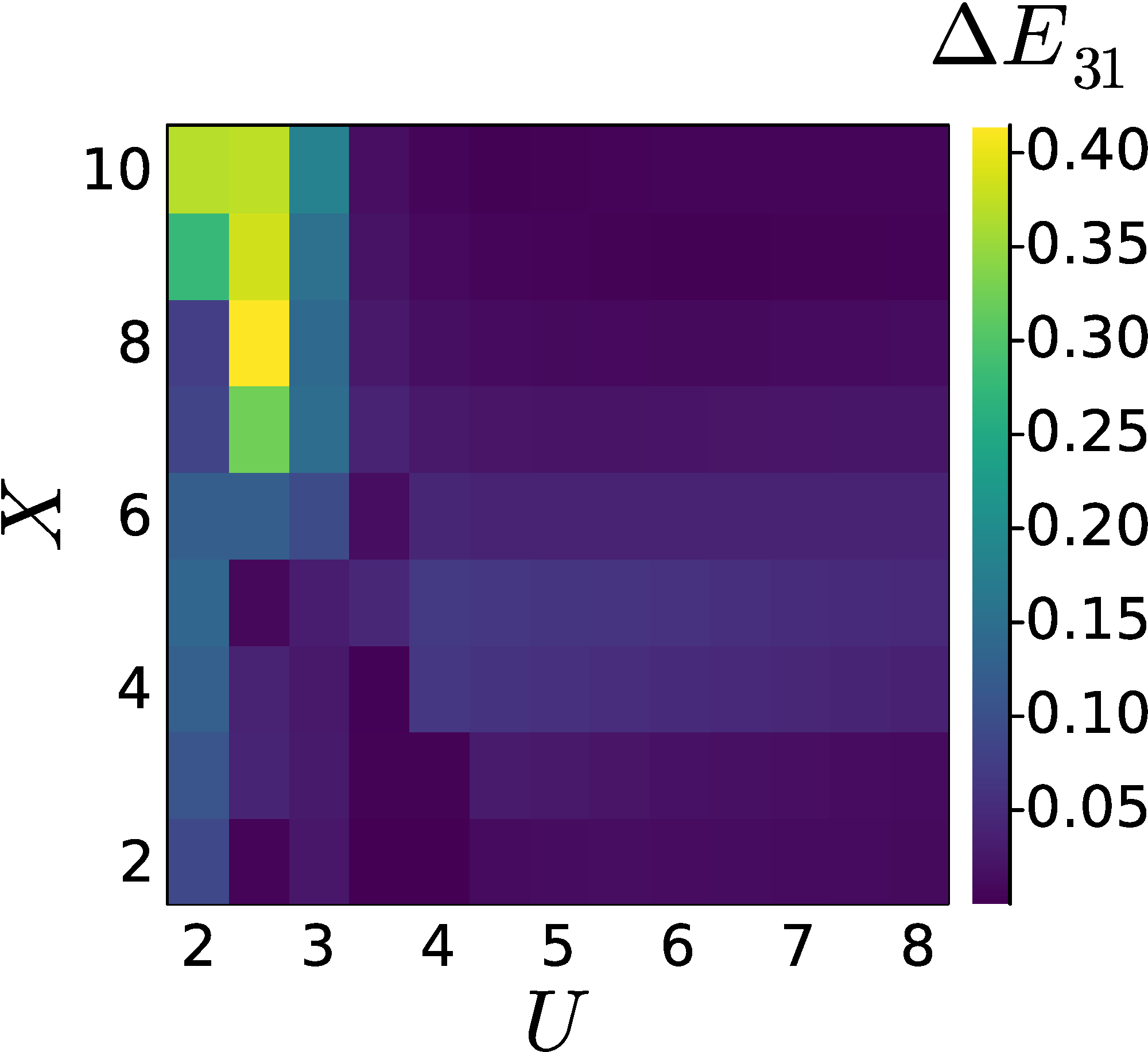}\\[2pt]
    \pnl{c}{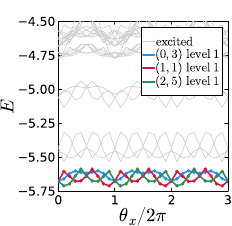}\hfill
    \pnl{d}{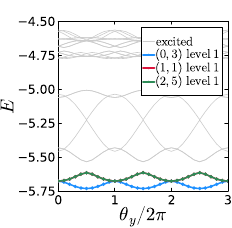}\\[2pt]
    \pnl{e}{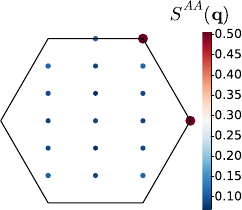}\hfill
    \pnl{f}{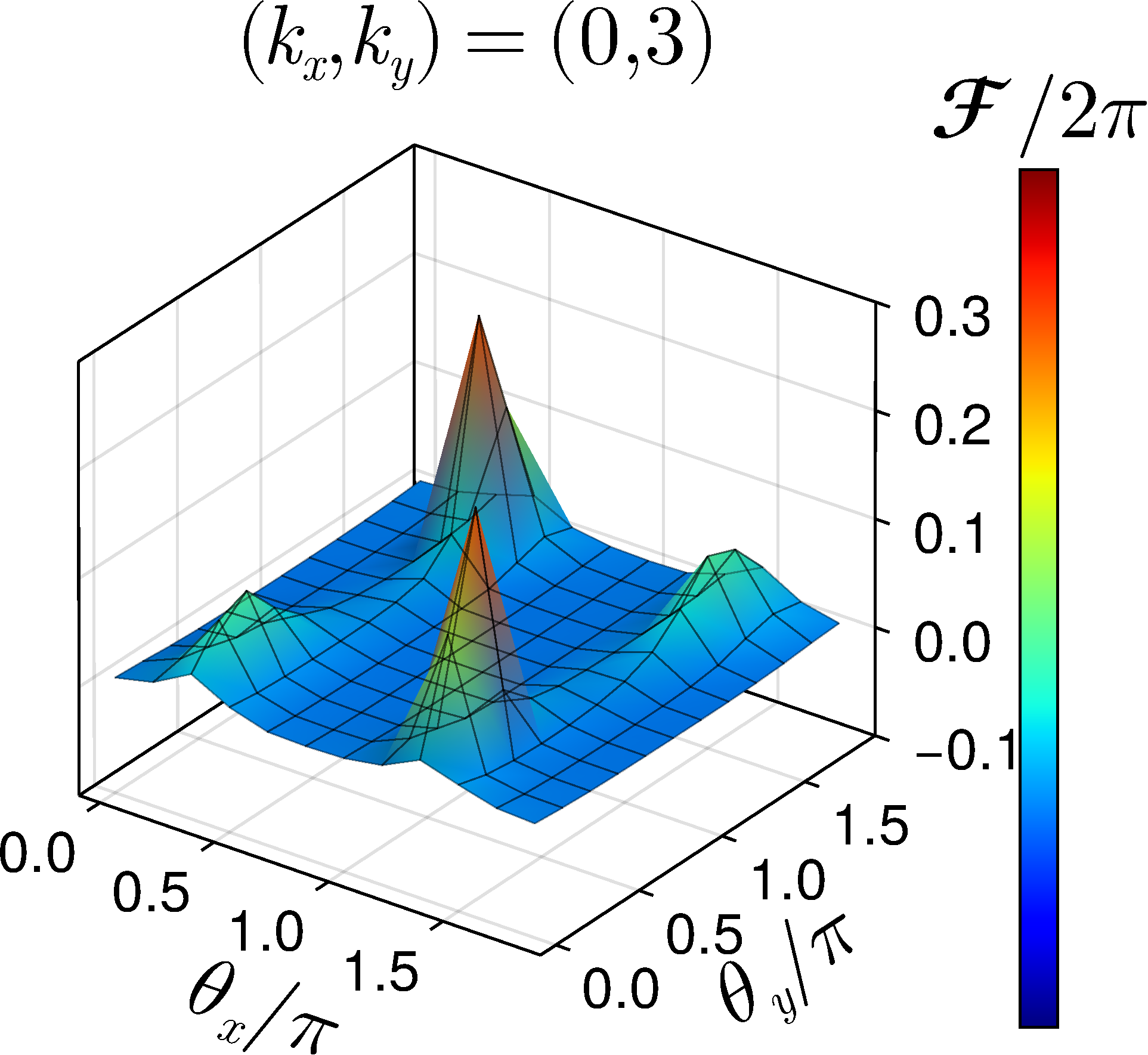}%
    \caption{ED results for the effective Hamiltonian for spin-polarized electrons on $N_s=36$ sites at filling $\nu=1/3$ ($N_f=6$). (a,b) Many-body gap $\Delta E_{43}$ and ground-state splitting $\Delta E_{31}$ against $(U, X)$. (c--f) Properties of the threefold ground-state manifold at $(U,X)=(4.5,10)$: (c,d) spectral flow of the ground-state energies against the twisted boundary phases $\theta_x$ and $\theta_y$; (e) intra-sublattice structure factor $S^{AA}(\mathbf{q})$; (f) Berry curvature $\mathcal{F}(\theta_x, \theta_y)$ for the ground state in momentum sector $(0,3)$ on a $12 \times 12$ grid.}
    \label{fig:fig4}
\end{figure}

Figure~\ref{fig:fig4} shows the ED results at filling $\nu=1/3$ for $N_s = 36$. From the many-body gap $\Delta E_{43}$ and the ground-state splitting $\Delta E_{31}$ in Figs.~\ref{fig:fig4}(a,b), a gapped phase with a threefold degenerate ground state appears in the region with $U>4$ and large $X$, while the rest of the diagram shows no comparable gap.
The structure factor [Fig.~\ref{fig:fig4}(e)] identifies this phase as a charge-density wave (CDW) with order at the $K$ points, whose three translates account for the threefold degeneracy~\cite{SupplementalMaterial}.
When we thread flux $\theta_x$ or $\theta_y$ [Figs.~\ref{fig:fig4}(c,d)], the spectral flow is $2\pi$-periodic and the three ground states return to themselves after one flux quantum rather than being permuted, so each state stays in its own momentum sector, where it carries a well-defined Chern number.
The many-body Chern number, computed from the Berry curvature at $(U,X)=(4.5,10)$ [Fig.~\ref{fig:fig4}(f)], is $C=2$ for the state in each sector, corresponding to a Hall conductance $\sigma_{xy}=2\,e^2/h$, while the parent Haldane band carries only $C=1$.
The doubled Chern number is consistent with band folding by the charge order: the $\sqrt{3}\times\sqrt{3}$ CDW at $K$ triples the unit cell and folds the $C=1$ Haldane band into three sub-bands, whose Chern numbers need only sum to $1$. In a mean-field model of the order, the lowest sub-band, filled at $\nu=1/3$, is gapped and carries $C=2$~\cite{SupplementalMaterial}.
These signatures show that strong electron-phonon coupling can stabilize a $C=2$ QAHC.

At $\nu=2/3$ we instead find a topologically trivial stripe CDW [End Matter, Fig.~\ref{fig:fig6}].

{\it Discussion.---}
Our results run counter to the common expectation that electron-phonon coupling destroys band topology. At partial filling of a Chern band, the coupling can instead stabilize a variety of topological phases: it generates the longer-range interactions needed to correlate the carriers, and in the attractive channel it also binds them into charge-$2e$ bosons. For spinful electrons this induces a bosonic FCI of bipolarons, flanked by superconductors at the $M$ and $K$ points and by several charge-ordered solids; for spin-polarized electrons, a $C=2$ QAHC at $\nu=1/3$. Notably, a flat band need not be prepared at the single-electron level: the flat Chern band hosting the bosonic FCI is an emergent property of the bipolarons, the underlying electron band being strongly dispersive, while for spin-polarized electrons the coupling compresses all kinetic scales, including the band gap, far below the generated interactions. The bipolaron FCI is stabilized by strong electron-phonon coupling and exhibits charge fractionalization: the constituent charge-$2e$ bipolarons fractionalize into emergent charge-$e$ semions. Although these quasiparticles carry the same charge as electrons, they obey distinct fractional quantum statistics.

The ingredients required for our mechanism---a topological band and strong electron-phonon coupling---are not exotic, and the interplay uncovered here may be particularly relevant to moir\'e transition-metal dichalcogenides and rhombohedral graphene, which are currently under intensive investigation as platforms for correlated topological states.
More broadly, our work suggests that electron-phonon coupling can act as a knob for engineering, rather than only destroying, topological matter.

{\it Acknowledgments.---}
We thank Xiaodong Hu for helpful discussions.
This work was facilitated through the use of advanced computational, storage, and networking infrastructure provided by the AI-core as well as the Hyak supercomputer system funded by the University of Washington Molecular Engineering Materials Center at the University of Washington (DMR-2308979). Exact diagonalization computations were performed using the XDiag library~\cite{Wietek2026:XDiag}. The work at LANL was supported by DOE via LDRD program at LANL and in part, by the Center for Integrated Nanotechnologies, an Office of Science User Facility operated for the U.S. DOE Office of Science, under user proposals 2018BU0010 and 2018BU0083. 
D.X. is supported by DOE Award No. DE-SC0012509. 
The manuscript writing by S.Z.L. was performed in part at Aspen Center for Physics, which is supported by National Science Foundation grant PHY-2210452.

\makeatletter%
\@ifundefined{auto@bib@empty}{}{\auto@bib@empty}%
\makeatother%

\section*{End Matter}

\begin{figure*}[t]
    \noindent
    \pnl{a}{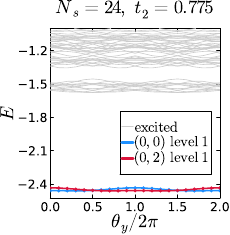}\hfill
    \pnl{b}{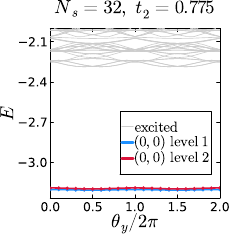}\hfill
    \pnl{c}{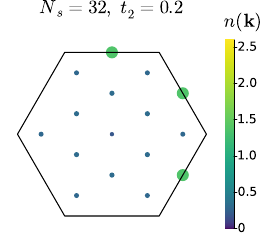}\hfill
    \pnl{d}{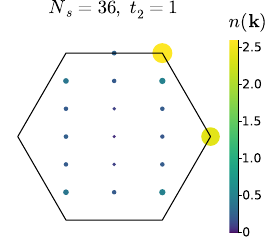}%
    \caption{ED results for the bipolaron model at $\kappa=1.5$. (a,b) Ground-state spectral flow for the FCI under $y$-direction flux insertion, with parameters as in Fig.~\ref{fig:fig1}, for (a) $N_s=24$ and (b) $N_s=32$. (c,d) Boson momentum occupation number $n(\mathbf{k})$ of the ground states, for (c) $t_2=0.2$ on $N_s=32$ and (d) $t_2=1$ on $N_s=36$.}
    \label{fig:fig5}
\end{figure*}

{\it Strong-coupling expansion.---}
The Lang--Firsov transformation~\cite{LangFirsov} $\hat{U}=\prod_{i} \exp[i(\alpha/ M \omega^2) \hat{p}_{i} \hat{n}_{i}] $ removes the electron-phonon coupling term of Eq.~(\ref{eq:spinful-orig}) in favor of phonon-dressed hoppings, giving
\begin{align}
    &\hat{U}^{\dagger} \hat{H} \hat{U} = - t_1 \sum_{\langle i,j \rangle, \sigma} \hat{S}_{ij} \hat{c}_{i \sigma}^{\dagger} \hat{c}_{j \sigma} 
    - t_{2} \sum_{\langle\langle i,j\rangle\rangle, \sigma} e^{i \nu_{ij}\phi} \hat{S}_{ij}  \hat{c}_{i \sigma}^{\dagger} \hat{c}_{j \sigma} \nonumber\\
    & - t_{3}\sum_{\langle\langle\langle i,j\rangle\rangle\rangle, \sigma}  e^{i \mu_{ij} \phi_2} \hat{S}_{ij} \hat{c}_{i \sigma}^{\dagger} \hat{c}_{j \sigma}
    + \frac{U_{\text{eff}}}{2} \sum_{i} \hat{n}_{i}^{2} 
    + \omega \sum_{i} \hat{a}_i^\dagger \hat{a}_i,
    \label{eq:spinful-transformed}
\end{align}
where $\hat{a}_i = \sqrt{M \omega/2}\, \hat{x}_i + i \hat{p}_i/\sqrt{2 M \omega}$ (we drop the zero-point energy $\omega/2$ per site), and $\hat{S}_{ij} = \hat S_i^\dagger \hat S_j = \exp[i \alpha (\hat{p}_j - \hat{p}_i)/( M \omega^2)]$ with
$\hat{S}_i = \exp[i \alpha \hat{p}_i/( M \omega^2)] = \exp[-\sqrt{X/2}\,(\hat{a}_i^\dagger - \hat{a}_i)]$ a phonon displacement operator, whence $\langle 0 |\hat{S}_{i}| 0 \rangle = e^{-X/4}$.

To second order in the dressed hopping terms, the effective Hamiltonian in the ground-state manifold of $\hat{H}_0$ is
\begin{equation}\label{eq:projector}
  \hat{H}_{\text{eff}} = \hat{P} \hat{H}_{t} \hat{P} + \hat{P} \hat{H}_{t} \hat{Q} \frac{1}{E_{0}-\hat{H}_{0}} \hat{Q} \hat{H}_{t} \hat{P},
\end{equation}
where
\begin{align}
  \hat{H}_{t} ={} &- t_1 \sum_{\langle i,j \rangle, \sigma} \hat{S}_{ij} \hat{c}_{i \sigma}^{\dagger} \hat{c}_{j \sigma}
  - t_{2} \sum_{\langle\langle i,j\rangle\rangle, \sigma} e^{i\nu_{ij}\phi} \hat{S}_{ij}  \hat{c}_{i \sigma}^{\dagger} \hat{c}_{j \sigma} \nonumber\\
  &- t_{3}\sum_{\langle\langle\langle i,j\rangle\rangle\rangle, \sigma} e^{i \mu_{ij} \phi_2} \hat{S}_{ij} \hat{c}_{i \sigma}^{\dagger} \hat{c}_{j \sigma},
\end{align}
$\hat{P}$ projects onto the ground-state manifold of $\hat{H}_0$, of energy $E_0$, and $\hat{Q}=\hat{I}-\hat{P}$ projects onto its orthogonal complement.

\begin{widetext}
For $U_{\text{eff}}<0$, evaluating Eq.~(\ref{eq:projector}) in the bipolaron ground-state manifold and integrating out the phonons gives
\begin{align}
    \frac{|U_{\text{eff}}|}{2 {F}(-X,Y) } \hat{H}_{\text{eff}} =& 
    - 
    \bigg[ 
    t_1^2 \sum_{\langle i,j \rangle } \hat{b}_i^\dagger \hat{b}_j  
    + t_2^2 \sum_{\langle\langle i,j \rangle\rangle } e^{2i\nu_{ij}\phi} \hat{b}_i^\dagger \hat{b}_j  
    + t_3^2 \sum_{\langle\langle\langle i,j\rangle\rangle\rangle}  e^{2i \mu_{ij}\phi_2} \hat{b}_i^\dagger \hat{b}_j  + \text{H.c.} 
    \bigg] \nonumber\\
    &+ \frac{2 {F}(X,Y) }{{F}(-X,Y)} \times 
    \bigg[
    t_1^2 \sum_{\langle i,j \rangle } \hat{b}_{i}^\dagger \hat{b}_i \hat{b}_j^\dagger \hat{b}_j  + t_2^2 \sum_{\langle\langle i,j \rangle\rangle } \hat{b}_{i}^\dagger \hat{b}_i \hat{b}_j^\dagger \hat{b}_j + t_3^2 \sum_{\langle\langle\langle i,j\rangle\rangle\rangle} \hat{b}_i^\dagger \hat{b}_i \hat{b}_j^\dagger \hat{b}_j \bigg] ,
    \label{eq:eff_ham}
\end{align}
with $F$, $X$ and $Y$ as defined in the main text.

For $U_{\text{eff}}>0$ in the spin-polarized limit, the same expansion on $\hat{H}_0 = \omega \sum_{i} \hat{a}_i^\dagger \hat{a}_i$ gives
\begin{align}\label{eq:polaron-second}
    e^{X/2}\, \hat{H}_{\text{eff}} =& -
    \Bigg[ t_1\sum_{\langle i,j \rangle} \hat{c}_{j}^{\dagger} \hat{c}_{i}
    +t_2 \sum_{\langle\langle i,j \rangle\rangle} e^{i\nu_{ij}\phi}
    \hat{c}_{i}^{\dagger} \hat{c}_{j}
    + \frac{2 F'(X/2)}{U_{\text{ph}}} \Bigg(  t_{1}^2  \sum_{\langle i,m,j \rangle}
    \hat{c}_{j}^{\dagger} (1-2\hat{n}_{m} ) \hat{c}_{i}
    + t_1 t_2 \sum_{\langle [ i,m,j ]\rangle}
    e^{i \nu_{jm}\phi}
    \hat{c}_{j}^{\dagger} (1-2\hat{n}_{m} ) \hat{c}_{i}  \nonumber\\
    &+  t_{2}^2  \sum_{\langle\langle i,m,j \rangle\rangle} e^{i (\nu_{jm}+\nu_{mi})\phi}
     \hat{c}_{j}^{\dagger} (1-2\hat{n}_{m} ) \hat{c}_{i} \Bigg) + \text{H.c.} \Bigg]
    + \frac{2 t_1^2 e^{X/2}}{U_{\text{ph}}} F'(X) \sum_{\langle i,j \rangle} \hat{n}_{i} \hat{n}_{j} + \frac{2 t_2^2 e^{X/2}}{U_{\text{ph}}} F'(X) \sum_{\langle \langle i,j \rangle\rangle} \hat{n}_{i} \hat{n}_{j},
\end{align}
\end{widetext}
where $\langle i,m,j \rangle$ denotes that $m$ is an NN of both $i$ and $j$, $\langle\langle i,m,j \rangle\rangle$ that $m$ is a 2NN of both, and $ \langle [ i,m,j ]\rangle$ that $m$ is an NN of $i$ and a 2NN of $j$.

The resulting effective Hamiltonian remains a valid perturbative expansion as long as $|U_\text{eff}| \gtrsim t_{1,2,3} e^{-X/2}$ and $\omega \gtrsim t_{1,2,3} e^{-X/2} \sqrt{X/2}$~\cite{Han2020:Strong}; the Supplemental Material~\cite{SupplementalMaterial} derives both these conditions and Eqs.~(\ref{eq:eff_ham}) and~(\ref{eq:polaron-second}).

{\it Further ED diagnostics of the bipolaron model.---}
Figure~\ref{fig:fig5} shows further diagnostics of the bipolaron model at $\kappa=1.5$. In the spectral flow under $y$-direction flux insertion at the maximal-flatness point $t_2=\sqrt{0.6}$ [Figs.~\ref{fig:fig5}(a,b)], the two FCI ground states exchange under each $2\pi$ of flux for $N_s=24$ and avoid crossing for $N_s=32$. The momentum occupation number $n(\mathbf{k})$ [Figs.~\ref{fig:fig5}(c,d)] shows the superfluids at $t_2=0.2$ and $t_2=1$ condensed at the $M$ and $K$ points, respectively.

\begin{figure*}[t]
    \noindent
    \pnl{a}{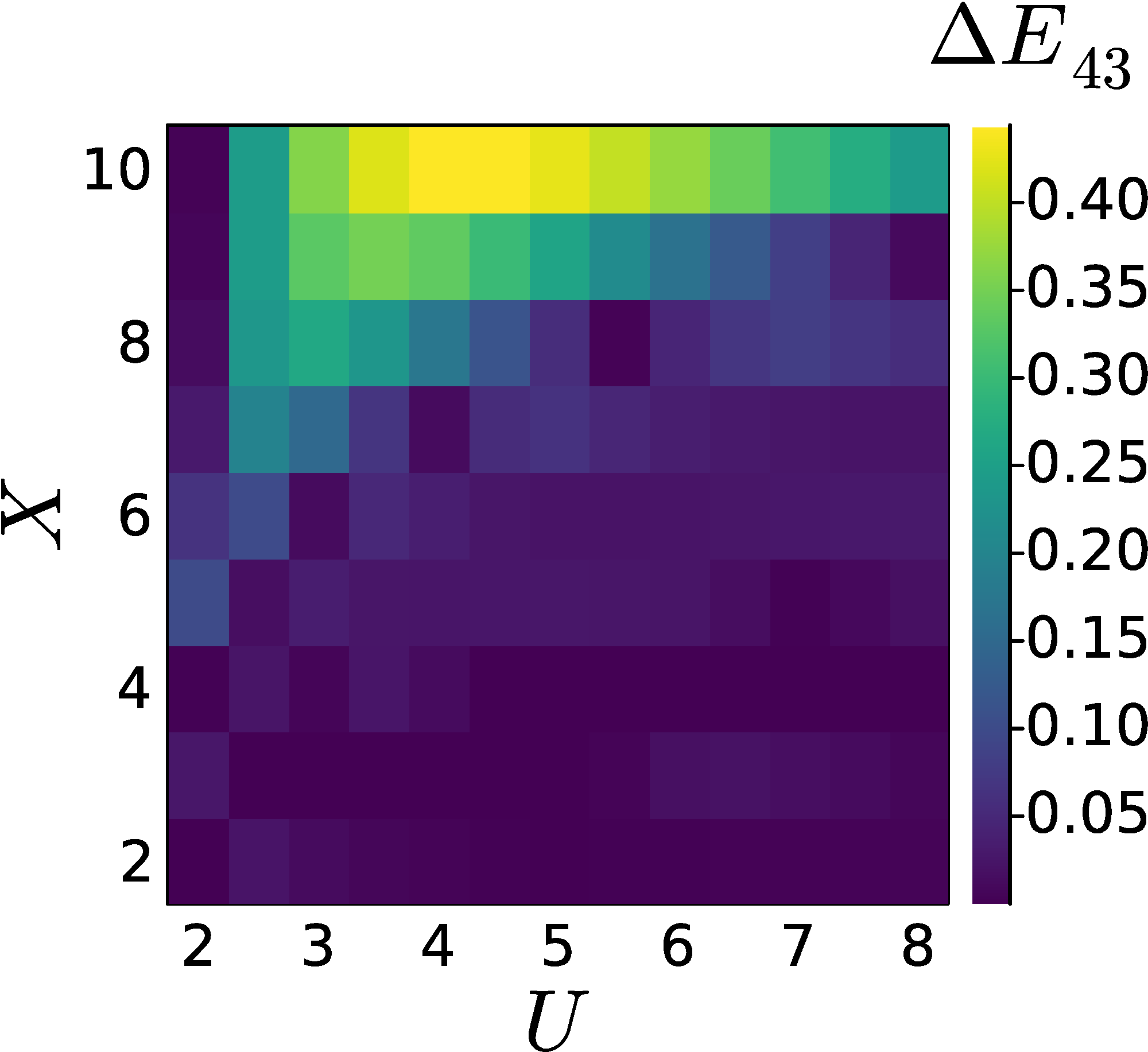}\hfill
    \pnl{b}{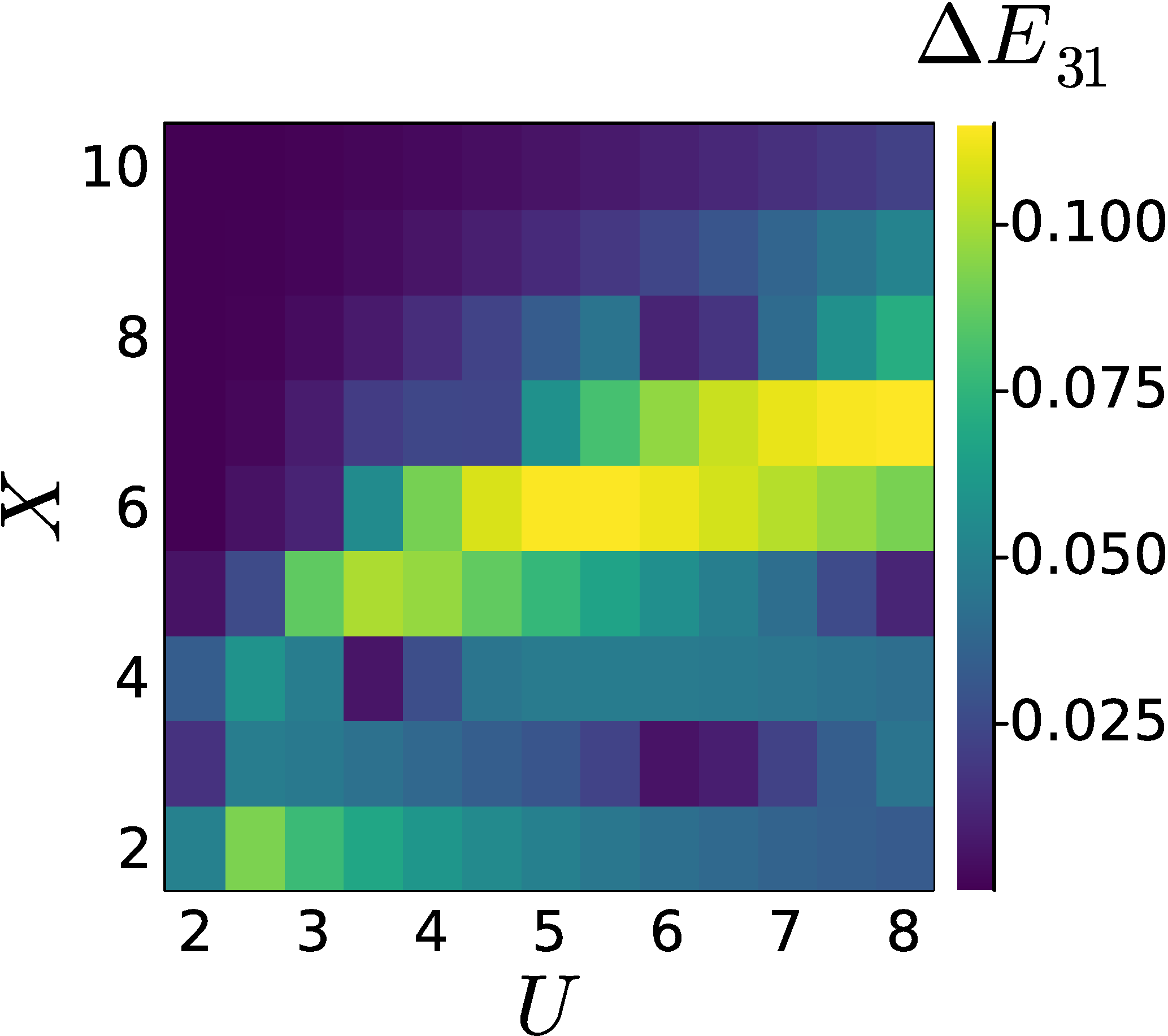}\hfill
    \pnl{c}{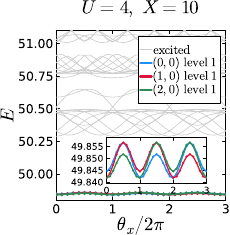}\hfill
    \pnl{d}{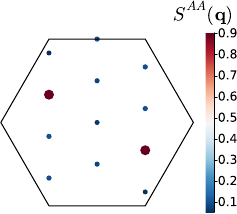}%
    \caption{ED results for the effective Hamiltonian for spin-polarized electrons on $N_s=24$ sites at filling $\nu=2/3$. The parameters $t_1$, $t_2$, and $\phi$ are chosen to maximize the flatness ratio of the bare Haldane band, as in Fig.~\ref{fig:fig4}. (a,b) Many-body gap $\Delta E_{43}$ and ground-state splitting $\Delta E_{31}$ against $(U, X)$. (c) Spectral flow against the twisted boundary phase $\theta_x$ at $(U,X)=(4,10)$. Inset: zoom on the ground-state manifold, whose three states are permuted by each $2\pi$ of flux. (d) Intra-sublattice structure factor $S^{AA}(\mathbf{q})$ at $(U,X)=(4,10)$.}
    \label{fig:fig6}
\end{figure*}

{\it Lattice conventions and observables.---}
We use the primitive lattice vectors $\mathbf{a}_1 = (\sqrt{3}, 0)$, $\mathbf{a}_2 = \left({\sqrt{3}}/{2}, {3}/{2}\right)$; the AB sublattices are at $\boldsymbol{\delta}_A = (0, 0)$, $\boldsymbol{\delta}_B=\left({\sqrt{3}}/{2}, {1}/{2}\right)$; the dual reciprocal lattice vectors, $\mathbf{a}_i\cdot\mathbf{b}_j = 2\pi\delta_{ij}$, are $\mathbf{b}_1 = \left({2\pi}/{\sqrt{3}}, -{2\pi}/{3}\right)$ and $\mathbf{b}_2 = \left(0, {4\pi}/{3}\right)$.
The ED energies quoted are eigenvalues of the rescaled Hamiltonians on the left-hand sides of Eqs.~(\ref{eq:eff_ham}) and~(\ref{eq:polaron-second}).
The superfluid stiffness is computed as
$\rho_s^{xx} = ({N_x}/{N_y})\, [E(+\delta\theta_x) + E(-\delta\theta_x) - 2E(0)]/(\delta\theta_x)^2$, where $E(\theta)$ is the ground-state energy under a twisted boundary phase $\theta$ and $\delta\theta_x = 0.01$.
The sublattice-resolved structure factor is
$S^{\alpha\beta}(\mathbf{q}) = (N_x N_y)^{-1} \sum_{i,j} e^{+i \mathbf{q} \cdot (\mathbf{R}_i - \mathbf{R}_j)} [\langle n_{i,\alpha} n_{j,\beta} \rangle - \langle n_{i,\alpha} \rangle \langle n_{j,\beta} \rangle]$, where $i,j \in [1, N_x N_y]$ are unit cell indices and $\alpha, \beta=A,B$ are sublattice indices.
The boson momentum occupation number is $n(\mathbf k)=(N_x N_y)^{-1}\sum_{i,j, \alpha}e^{i\mathbf k\cdot(\mathbf R_i-\mathbf R_j)}\, \langle b_{i,\alpha}^\dagger\,b_{j,\alpha}\rangle$.
Under twisted boundary phases $(\theta_x, \theta_y)$, the Berry curvature of a many-body state $|\psi\rangle$ is
$\mathcal{F}(\theta_x,\theta_y) = -2\operatorname{Im}\langle \partial_{\theta_x}\psi \,|\, \partial_{\theta_y}\psi\rangle$
and its Chern number is $C=(2\pi)^{-1}\iint \mathcal{F}(\theta_x,\theta_y)\,d\theta_x\,d\theta_y$.

{\it Limiting cases of the spin-polarized model.---}
Two limiting cases of the bare phonon frequency $\omega$, at fixed coupling strength $\alpha$, are instructive.
In the antiadiabatic limit $\omega / t_1 \to \infty$,
the model reduces exactly to the bare Haldane model, as no phonon excitations are accessible at low energy.
In the adiabatic limit $\omega/ t_1 \to 0$, the hoppings are suppressed exponentially faster than the interactions, so any ground-state order survives only at vanishingly small temperatures.

{\it Stripe charge order at $\nu=2/3$.---}
Figure~\ref{fig:fig6} shows the numerical results for filling $\nu=2/3$ and $N_s=24$. In roughly the same parameter range where the $\nu=1/3$ QAHC appears [Figs.~\ref{fig:fig6}(a,b)], we find a stripe CDW phase, manifested by the prominent structure factor peaks at momentum $ \pm \mathbf{b}_{1}/3$ in Fig.~\ref{fig:fig6}(d). In contrast to the QAHC, the three ground states are linked by flux insertion: every $2\pi$ flux threaded in the $x$ direction permutes them among one another [Fig.~\ref{fig:fig6}(c)] rather than leaving each pinned to a fixed sector. Their total many-body Chern number is $C_{\text{tot}}=0$, confirming a trivial topology.

\end{document}

\title{Supplemental Material for ``Haldane--Holstein model at fractional filling: Route to bosonic fractional Chern insulator and quantum anomalous Hall crystal''}
\author{Zezhu Wei}
\affiliation{Department of Materials Science and Engineering, University of Washington, Seattle, Washington 98195, USA}

\author{Ang-Kun Wu}
\affiliation{Department of Physics and Astronomy, University of Tennessee, Knoxville, Tennessee 37996, USA}

\author{Di Xiao}
\affiliation{Department of Materials Science and Engineering, University of Washington, Seattle, Washington 98195, USA}
\affiliation{Department of Physics, University of Washington, Seattle, Washington 98195, USA}

\author{Shi-Zeng Lin}
\affiliation{Theoretical Division, T-4 and CNLS, Los Alamos National Laboratory, Los Alamos, New Mexico 87545, USA}
\affiliation{Center for Integrated Nanotechnologies (CINT), Los Alamos National Laboratory, Los Alamos, New Mexico 87545, USA}
\maketitle

\tableofcontents

\section{Bosonic effective Hamiltonian: $U_{\text{eff}}<0$}
\label{sec:sm-boson}

This and the next section parallel the strong-coupling expansion of Ref.~\cite{Han2020:Strong} for the Holstein--Hubbard model; we go beyond to include the second- and third-nearest-neighbor (2NN and 3NN) hopping channels of the extended Haldane model, tracking their complex phases through the expansion. We start from the Lang--Firsov-transformed Hamiltonian, split as $\hat{H} = \hat{H}_0 + \hat{H}_t$ with
\begin{align}
    \hat{H}_{0} &= \frac{U_{\text{eff}}}{2} \sum_{i} \hat{n}_{i}^{2}
    + \omega \sum_{i} \hat{a}_{i}^{\dagger} \hat{a}_{i},
    \\
    \hat{H}_{t} &= - t_1 \sum_{\langle i,j \rangle, \sigma} \hat{S}_{ij} \hat{c}_{i \sigma}^{\dagger} \hat{c}_{j \sigma}
    - t_{2} \sum_{\langle\langle i,j\rangle\rangle, \sigma} e^{i\nu_{ij}\phi} \hat{S}_{ij}  \hat{c}_{i \sigma}^{\dagger} \hat{c}_{j \sigma}
    \nonumber\\
    &\quad - t_{3} \sum_{\langle\langle\langle i,j\rangle\rangle\rangle, \sigma} e^{i\mu_{ij}\phi_2} \hat{S}_{ij}  \hat{c}_{i \sigma}^{\dagger} \hat{c}_{j \sigma},
\end{align}
where $\hat{S}_{ij} = \exp[i\alpha(\hat{p}_j - \hat{p}_i)/(M\omega^2)]$ and constant terms are dropped. When $U_{\text{eff}}<0$ and $|U_{\text{eff}}|$ dominates over the hoppings, the zeroth-order ground-state manifold consists of states in which every site is either empty or occupied by an on-site bipolaron, $\hat{b}_{i}^{\dagger} = \hat{c}_{i\uparrow}^{\dagger} \hat{c}_{i\downarrow}^{\dagger}$, with no phonon excitations. To second order in $\hat{H}_t$, degenerate perturbation theory gives~\cite{Han2020:Strong}
\begin{equation}
    \hat{H}_{\text{eff}} = \hat{P} \hat{H}_{t} \hat{P} + \hat{P} \hat{H}_{t} \hat{Q} \frac{1}{E_{0}-\hat{H}_{0}} \hat{Q} \hat{H}_{t} \hat{P},
\end{equation}
where $\hat{P}$ projects onto the ground-state manifold, of energy $E_0$, and $\hat{Q}=\hat{I}-\hat{P}$. The first-order term vanishes, since a single hopping process leaves the ground-state manifold. The second-order term is
\begin{equation}
    \hat{H}_{\text{eff}} = \sum_{\alpha \beta b} |\alpha\rangle \langle \alpha |\hat{H}_{t} |b\rangle \frac{1}{E_{0}-E_{b}} \langle b | \hat{H}_{t} |\beta\rangle \langle \beta|,
    \label{eq:sm-projector}
\end{equation}
where $|\alpha\rangle$ and $|\beta \rangle$ are ground states of $\hat{H}_0$, and $|b\rangle$ is an excited state with $\hat{H}_0 |b\rangle = E_{b} |b\rangle$. Two distinct virtual processes arise from the nearest-neighbor (NN) hopping.

{\it Case I (density-density interaction).---}
\begin{align}
    |\beta \rangle &= \hat{c}_{i \uparrow}^{\dagger} \hat{c}_{i \downarrow}^{\dagger} |\phi\rangle
    \equiv | {\uparrow\downarrow},0;0,0\rangle,
    \\
    |b \rangle &= \hat{c}_{i \sigma}^{\dagger} \hat{c}_{j \bar{\sigma}}^{\dagger} \frac{(\hat{a}_{i}^\dagger)^n}{\sqrt{n!}}\frac{(\hat{a}_{j}^\dagger)^\ell}{\sqrt{\ell!}} |\phi\rangle
    \equiv | \sigma,\bar{\sigma};n,\ell\rangle,
    \\
    |\alpha\rangle &= \hat{c}_{i \uparrow}^{\dagger} \hat{c}_{i \downarrow}^{\dagger} |\phi\rangle
    \equiv | {\uparrow\downarrow},0;0,0\rangle,
\end{align}
where $i,j$ are NN sites, $|\phi \rangle$ is an arbitrary ground state with $i$ and $j$ empty, and the kets $| \cdot, \cdot;\cdot,\cdot \rangle$ list the electron occupations of sites $i,j$ followed by their phonon numbers. With $E_b-E_0=|U_{\text{eff}}|+(n+\ell)\omega$, and with the spin sum and the process with $i$ and $j$ interchanged (the bipolaron on site $j$) each yielding a factor of $2$,
\begin{align}
    \hat{H}_{\text{eff}}^{(\text{I})}
    ={}& - 4\sum_{n,\ell} \langle 0,0 | \hat{S}_{ij} |n,\ell\rangle  \langle n,\ell | \hat{S}_{ji} |0,0\rangle \nonumber\\
    &\times \frac{t_{1}^2}{|U_{\text{eff}}|+(n+\ell)\, \omega}
    \sum_{\langle i,j \rangle} \hat{c}_{i \uparrow}^{\dagger} \hat{c}_{i \downarrow}^{\dagger} |\phi\rangle \langle\phi| \hat{c}_{i \downarrow} \hat{c}_{i \uparrow}.
    \label{eq:sm-caseI}
\end{align}
The phonon overlaps follow from the displaced-oscillator matrix element
\begin{align}
    \langle 0 | e^{\pm i\alpha \hat{p}/(M\omega^2)} |n \rangle = e^{-X/4} \left(\pm \sqrt{\frac{X}{2}}\right)^n \frac{1}{\sqrt{n!}},
    \label{eq:sm-fc-overlap}
\end{align}
where $X = U_{\text{ph}}/\omega = \alpha^2/(M\omega^3) > 0$, so that $\langle 0,0 | \hat{S}_{ij} |n,\ell\rangle \langle n,\ell | \hat{S}_{ji} |0,0\rangle = e^{-X} (X/2)^{n+\ell}/(n!\,\ell!)$. Within the ground-state manifold, $\hat{c}_{i \uparrow}^{\dagger} \hat{c}_{i \downarrow}^{\dagger} |\phi\rangle \langle\phi| \hat{c}_{i \downarrow} \hat{c}_{i \uparrow} = \hat{b}_{i}^\dagger \hat{b}_i\, (1 - \hat{b}_j^\dagger \hat{b}_j)$; the first term counts the boson number and is a constant at fixed filling. Dropping it,
\begin{align}
    \hat{H}_{\text{eff}}^{(\text{I})}
    = \frac{4t_1^2}{|U_{\text{eff}}|}\, F(X,Y)
    \sum_{\langle i,j \rangle } \hat{b}_{i}^\dagger \hat{b}_i\, \hat{b}_j^\dagger \hat{b}_j,
\end{align}
where $Y=|U_{\text{eff}}|/\omega > 0$ and $F$ is defined~\cite{Han2020:Strong}, for either sign of its first argument and for $y>0$, by
\begin{align}
    F(x,y) &= \sum_{n,\ell} \frac{y\, e^{-|x|}}{y+ (n+\ell) }  \left(\frac{x}{2}\right)^{n+\ell} \frac{1}{n!\, \ell!}
    \nonumber\\
    &= y\, e^{-|x|} \int_{0}^{1} dz \, z^{y-1} e^{x z}
    \nonumber\\
    &= e^{-|x|}\, {}_{1}F_{1}(y;\, y+1;\, x),
    \label{eq:sm-F}
\end{align}
evaluated at $(x,y) = (X,Y)$ in Case I.  The last line is the Euler integral representation of the confluent hypergeometric function, ${}_{1}F_{1}(a;b;x) = [\Gamma(b)/\Gamma(a)\Gamma(b-a)]\int_{0}^{1} dz\, e^{xz} z^{a-1}(1-z)^{b-a-1}$, at $a=y$, $b=y+1$~\cite[3.383.1]{Gradshteyn2007:Table}. 

{\it Case II (bipolaron hopping).---}Keeping $|\beta\rangle$ and $|b\rangle$ of Case I but taking
\begin{align}
    |\alpha\rangle = \hat{c}_{j \uparrow}^{\dagger} \hat{c}_{j \downarrow}^{\dagger} |\phi\rangle
    \equiv | 0,{\uparrow\downarrow};0,0\rangle,
\end{align}
both electrons hop $i \to j$, and each phonon overlap now carries the same sign, giving $(-X/2)^{n+\ell}$ in place of $(X/2)^{n+\ell}$, i.e., Eq.~(\ref{eq:sm-F}) evaluated at $x = -X$:
\begin{align}
    \hat{H}_{\text{eff}}^{(\text{II})}
    = -\frac{2t_1^2}{|U_{\text{eff}}|}\, F(-X,Y)
    \sum_{\langle i,j \rangle } (\hat{b}_i^\dagger \hat{b}_j + \text{H.c.}).
\end{align}
Because the prefactor in Eq.~(\ref{eq:sm-F}) is $e^{-|x|}$, both channels are suppressed by the same $e^{-X}$; explicitly, $F(-X,Y) = Y e^{-X} \int_{0}^{1} dz \, z^{Y-1} e^{-X z}$. For $X, Y>0$ this implies $0 < F(-X,Y) < F(X,Y)$: the bipolaron hopping is always weaker than the interaction, so the tuning parameter of the main text obeys $\kappa = F(X,Y)/F(-X,Y) > 1$.

{\it 2NN and 3NN channels.---}The processes built from the 2NN and 3NN hoppings repeat Cases I and II with $t_1 \to t_{2,3}$. In the interaction channel the complex phases of the two conjugate hops cancel, while in the hopping channel both electrons traverse the same oriented bond, so the phase doubles: $e^{i\nu_{ij}\phi} \to e^{2i\nu_{ij}\phi}$ and $e^{i\mu_{ij}\phi_2} \to e^{2i\mu_{ij}\phi_2}$. Collecting all channels yields Eq.~(5) of the main text.

\section{Effective Hamiltonian for spin-polarized electrons}
\label{sec:sm-polarized}

For spin-polarized electrons no site can be doubly occupied, so $\hat{n}_i^2 = \hat{n}_i$ and the on-site interaction is a constant at fixed filling; the unperturbed Hamiltonian is then $\hat{H}_0 = \omega \sum_{i} \hat{a}_{i}^{\dagger} \hat{a}_{i}$, whose ground-state manifold consists of states with arbitrary electron configurations and no phonon excitations. At first order the hopping is dressed by the Franck--Condon factor $\langle 0,0|\hat{S}_{ij}|0,0\rangle = e^{-X/2}$,
\begin{equation}
    \hat{H}_{\text{eff}}^{(1)} = -t_1 e^{-X/2} \sum_{\langle i,j \rangle} \hat{c}_{j}^{\dagger} \hat{c}_{i} + \text{H.c.}
\end{equation}
At second order, Eq.~(\ref{eq:sm-projector}) generates three virtual processes, which we derive explicitly for the NN hopping $t_1$.

{\it Case I (density-density interaction).---}
\begin{align}
    |\beta \rangle &= \hat{c}_{i }^{\dagger} |\phi\rangle
    \equiv | 1,0;0,0\rangle,
    \\
    |b \rangle &= \hat{c}_{j }^{\dagger} \frac{(\hat{a}_{i}^\dagger)^n}{\sqrt{n!}}\frac{(\hat{a}_{j}^\dagger)^\ell}{\sqrt{\ell!}} |\phi\rangle
    \equiv | 0,1;n,\ell\rangle,
    \\
    |\alpha\rangle &= \hat{c}_{i }^{\dagger} |\phi\rangle
    \equiv | 1,0;0,0\rangle,
\end{align}
with $n+\ell\neq 0$. Including the process with $i$ and $j$ interchanged (the electron on site $j$), which doubles the result, the second-order term is
\begin{align}
    \hat{H}_{\text{eff}}^{(\text{I})}
    ={}& -2\sum_{\substack{n,\ell\\ n+\ell \neq 0}} \langle 0,0|\hat{S}_{ij} |n,\ell\rangle  \langle n,\ell | \hat{S}_{ji} |0,0\rangle
    \nonumber\\
    &\times \frac{t_{1}^2}{(n+\ell)\, \omega} \sum_{\langle i,j \rangle}
    \hat{c}_{i }^{\dagger}  |\phi\rangle \langle \phi|
    \hat{c}_{i }.
\end{align}
Within the ground-state manifold, $\hat{c}_{i }^{\dagger}  |\phi\rangle \langle \phi| \hat{c}_{i } = \hat{c}_{i }^{\dagger} (1-\hat{n}_{j} ) \hat{c}_{i } \sim \hat{n}_{i} (1-\hat{n}_{j})$, and the term linear in $\hat{n}_i$ is a constant at fixed filling. Dropping it and using the definition
\begin{align}
    F'(x) &= \sum_{\substack{n,\ell\\ n+\ell \neq 0}} \frac{ x }{ n+\ell } e^{-x} \left(\frac{x}{2}\right)^{n+\ell} \frac{1}{n!\, \ell!}
    \nonumber\\
    &= x\, e^{-x} \int_{0}^{1} \frac{dz}{z}\, \bigl( e^{x z} - 1 \bigr)
    \nonumber\\
    &= x\, e^{-x} \bigl[ \operatorname{Ei}(x) - \ln x - \gamma \bigr],
    \label{eq:sm-Fp}
\end{align}
with $\operatorname{Ei}$ the exponential integral and $\gamma$ Euler's constant; the substitution $t = xz$ turns the second line into $\int_{0}^{x} dt\, (e^{t}-1)/t$, and the third is then the standard exponential-integral representation~\cite[8.212.1 and 8.214.2]{Gradshteyn2007:Table}.  
We arrive at
\begin{equation}
    \hat{H}_{\text{eff}}^{(\text{I})} = \frac{2t_1^2}{U_{\text{ph}}} F'(X) \sum_{\langle i,j \rangle} \hat{n}_{i} \hat{n}_{j}.
\end{equation}

{\it Case II (2NN hopping, empty bridge).---}Here $m$ denotes the bridging site, an NN of both $i$ and $j$:
\begin{align}
    |\beta \rangle &= \hat{c}_{i}^{\dagger}  |\phi\rangle
    \equiv |1,0,0;0,0,0\rangle,
    \\
    |b \rangle &= \hat{c}_{m}^{\dagger}   \frac{(\hat{a}_{m}^\dagger)^n}{\sqrt{n!}} |\phi\rangle
    \equiv |0,1,0;0,n,0\rangle,
    \\
    |\alpha\rangle &= \hat{c}_{j}^{\dagger} |\phi\rangle
    \equiv |0,0,1;0,0,0\rangle.
\end{align}
Since all prefactors are real, the reversed path $j \to m \to i$ gives the Hermitian conjugate, and the second-order term is
\begin{align}
    \hat{H}_{\text{eff}}^{(\text{II})}
    ={}& -\sum_{n\neq 0} \langle 0,0,0|\hat{S}_{jm} |0,n,0\rangle  \langle 0,n,0 | \hat{S}_{mi} |0,0,0\rangle
    \nonumber\\
    &\times \frac{t_{1}^2}{n\, \omega} \sum_{\langle i,m,j \rangle}
    \hat{c}_{j }^{\dagger}  |\phi\rangle \langle \phi|
    \hat{c}_{i }  + \text{H.c.}
\end{align}
Using
\begin{align}
    \sum_{n\neq 0} \frac{X}{2 n} \langle 0,0,0|\hat{S}_{jm} |0,n,0\rangle & \langle 0,n,0 | \hat{S}_{mi} |0,0,0\rangle
    \nonumber\\
    &= e^{-X/2} F'\!\left(\frac{X}{2}\right)
\end{align}
and $\hat{c}_{j }^{\dagger}  |\phi\rangle \langle \phi| \hat{c}_{i } = \hat{c}_{j }^{\dagger} (1- \hat{n}_{m} ) \hat{c}_{i }$, we obtain
\begin{align}
    \hat{H}_{\text{eff}}^{(\text{II})} ={}& -
    \frac{2 t_{1}^2}{U_{\text{ph}}} e^{-X/2} F'\!\left(\frac{X}{2}\right)
    \nonumber\\
    &\times \sum_{\langle i,m,j \rangle}
    \hat{c}_{j }^{\dagger} (1- \hat{n}_{m} ) \hat{c}_{i }  + \text{H.c.}
\end{align}

{\it Case III (2NN hopping, occupied bridge).---}The same 2NN process with the bridge occupied:
\begin{align}
    |\beta \rangle &= \hat{c}_{i }^{\dagger} \hat{c}_{m }^\dagger |\phi\rangle
    \equiv | 1,1,0;0,0,0\rangle,
    \\
    |b \rangle &= \hat{c}_{i}^{\dagger}  \hat{c}_{j}^\dagger \frac{(\hat{a}_{m}^\dagger)^n}{\sqrt{n!}} |\phi\rangle
    \equiv | 1,0,1;0,n,0\rangle,
    \\
    |\alpha\rangle &= \hat{c}_{m}^{\dagger} \hat{c}_{j}^\dagger |\phi\rangle
    \equiv | 0,1,1;0,0,0\rangle.
\end{align}
The projected operator reduces to
\begin{align}
    \hat{c}_{m }^{\dagger} \hat{c}_{j }^\dagger  |\phi\rangle \langle \phi|
    \hat{c}_{m } \hat{c}_{i }
    = \hat{c}_{m }^{\dagger} \hat{c}_{j }^\dagger
    \hat{c}_{m } \hat{c}_{i }
    = - \hat{c}_{j }^{\dagger} \hat{n}_{m} \hat{c}_{i },
\end{align}
so this process contributes
\begin{align}
    \hat{H}_{\text{eff}}^{(\text{III})} = \frac{2 t_{1}^2}{U_{\text{ph}}} e^{-X/2} F'\!\left(\frac{X}{2}\right) \sum_{\langle i,m,j \rangle}
    \hat{c}_{j }^{\dagger} \hat{n}_{m}  \hat{c}_{i } + \text{H.c.}
\end{align}

Cases II and III combine into the density-assisted hopping of the main text, the bridge occupation entering as $1-2\hat{n}_m$. Adding the first- and second-order terms, the total effective Hamiltonian is
\begin{align}
    \hat{H}_{\text{eff}} ={}& \bigg[ -t_1 e^{-X/2} \sum_{\langle i,j \rangle}
    \hat{c}_{j }^{\dagger} \hat{c}_{i }
    \nonumber\\
    &-
    \frac{2 t_{1}^2}{U_{\text{ph}}} e^{-X/2} F'(X/2) \sum_{\langle i,m,j \rangle}
    \hat{c}_{j}^{\dagger} (1-2 \hat{n}_{m} ) \hat{c}_{i}
    \nonumber\\
    &+ \text{H.c.} \bigg]
    + \frac{2t_1^2}{U_{\text{ph}}} F'(X) \sum_{\langle i,j \rangle} \hat{n}_{i} \hat{n}_{j}.
\end{align}
Taking one or both of the two virtual hops to be the 2NN hopping $t_2$ gives the remaining channels in the same way: the mixed $t_1 t_2$ process, in which $m$ is an NN of $i$ and a 2NN of $j$; the pure $t_2^2$ process, in which $m$ is a 2NN of both; and the 2NN density-density interaction of Case I. Each hop carries the complex phase of the link it traverses, so the mixed channel acquires the single factor $e^{i\nu_{jm}\phi}$ and the pure channel the product $e^{i(\nu_{jm}+\nu_{mi})\phi}$, while in the density-density channel the two conjugate hops lie on the same link and their phases cancel. Collecting these with the first-order hoppings, which include the Franck--Condon-dressed 2NN term, yields Eq.~(6) of the main text.

\begin{figure}[tbp]
    \centering
    \includegraphics{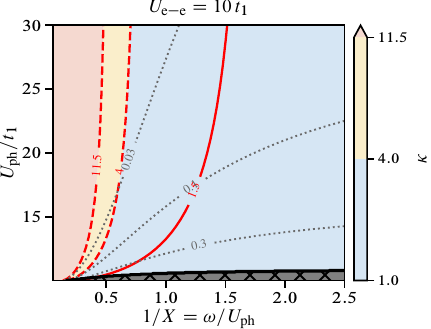}
    \caption{Parameter map of the effective hard-core boson model at maximal band flatness, at $U_{\text{e-e}}=10\,t_1$. Fill: $\kappa$, in the ranges that define the FCI ($\kappa<4$), Solid I ($4<\kappa<11.5$) and Solid II ($\kappa>11.5$). Red: the FCI point $\kappa=1.5$ (solid) and the two transitions (dashed). Gray dotted: the factor $2F(-X,Y)\,t_1^2/|U_{\text{eff}}|$ converting the exact-diagonalization (ED) energies of Eq.~(5) of the main text into bare units. Hatched region is where Eq.~(\ref{eq:sm-validity}) is violated, dominated by the pair-breaking bound $|U_{\text{eff}}| \lesssim t_1 e^{-X/2}$.}
    \label{fig:sm-window-boson}
\end{figure}

\begin{figure}[tbp]
    \centering
    \includegraphics{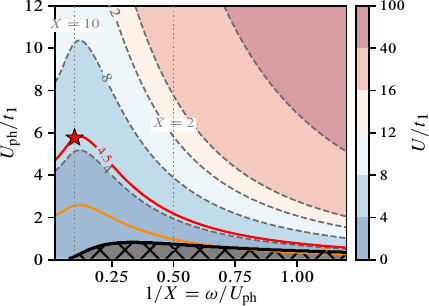}
    \caption{Parameter map of the spin-polarized model, filled by the parameter $U/t_1=U_{\text{ph}}/[F'(X/2)\,t_1]$ of the main text; constant $X$ is a vertical line (dotted lines at $X=2$, $10$). Red: $U=4.5\,t_1$; orange: $U=2\,t_1$, the lower edge of the $(U,X)$ scans in ED; star: the ED working point $(U,X)=(4.5,10)$ of the $\nu=1/3$ quantum anomalous Hall crystal. Hatched region is where the no-phonon condition of Eq.~(\ref{eq:sm-validity}) is violated.}
    \label{fig:sm-window-polaron}
\end{figure}

\section{Range of validity of the effective Hamiltonians}
\label{sec:sm-validity}

Following the analysis of Ref.~\cite{Han2020:Strong}, the effective Hamiltonians derived above operate in reduced Hilbert spaces with restricted site occupancies, determined by the sign of $U_{\text{eff}}$, and with zero phonon excitations: for $U_{\text{eff}}<0$ every site is either empty or occupied by a charge-$2e$ bipolaron, while for the spin-polarized model the restriction is to the phonon vacuum alone. These restrictions become invalid when excitation energies of the unperturbed Hamiltonian $\hat{H}_0$ are no longer large compared with the corresponding matrix elements of the transformed hopping.

The relevant matrix elements follow from Eq.~(\ref{eq:sm-fc-overlap}). Between zero-phonon states the hopping on a bond is suppressed by the Franck--Condon factor, $\langle 0,0|\hat{S}_{ij}|0,0\rangle = e^{-X/2}$, and the largest phonon-emitting element is $|\langle 1,0|\hat{S}_{ij}|0,0\rangle| = e^{-X/2}\sqrt{X/2}$. In the bosonic problem, breaking a bipolaron costs $|U_{\text{eff}}|$, and creating a phonon costs $\omega$. The restriction on site occupancies must therefore be lifted when $|U_{\text{eff}}| \lesssim t_{a}\, e^{-X/2}$, and phonon excitations must be included when $\omega \lesssim t_{a}\, e^{-X/2}\sqrt{X/2}$, where $t_{a}$ runs over the hoppings of the model. The effective Hamiltonians are therefore valid perturbative expansions as long as
\begin{align}
    |U_{\text{eff}}| \gtrsim t_{1,2,3}\, e^{-X/2},
    \qquad
    \omega \gtrsim t_{1,2,3}\, e^{-X/2}\sqrt{\frac{X}{2}},
    \label{eq:sm-validity}
\end{align}
the conditions quoted in the End Matter of the main text, with the $t_1$ conditions the tightest, since $t_1$ is the largest hopping. For the spin-polarized model only the phonon condition applies, since $\hat{H}_0=\omega\sum_i \hat{a}_i^\dagger \hat{a}_i$; in the variables $(U,X)$ of the main text, with $U = U_{\text{ph}}/F'(X/2)$ and hence $\omega = U F'(X/2)/X$, it reads
\begin{align}
    \frac{U}{t_1} \gtrsim \frac{X}{F'(X/2)}\, e^{-X/2}\sqrt{\frac{X}{2}}.
    \label{eq:sm-validity-UX}
\end{align}

Figures~\ref{fig:sm-window-boson} and \ref{fig:sm-window-polaron} locate the phases reported in the main text in parameter space, on the $(\omega/U_{\text{ph}},\,U_{\text{ph}}/t_1)$ plane, whose horizontal axis is $1/X$; the invalid regions are hatched.

For the bosonic model (Fig.~\ref{fig:sm-window-boson}), shown at fixed Hubbard repulsion $U_{\text{e-e}}=10\,t_1$, the plane divides into fractional Chern insulator (FCI), Solid I, and Solid II regions according to $\kappa$ alone: the FCI point $\kappa=1.5$ and the phase boundaries $\kappa=4$ and $11.5$ all lie inside the valid window, which fails only in the pair-breaking region $U_{\text{ph}}-U_{\text{e-e}}\lesssim t_1 e^{-X/2}$ along the bottom edge.

For the spin-polarized model (Fig.~\ref{fig:sm-window-polaron}), constant $X$ is a vertical line, and the orange contour $U=2\,t_1$ marks the lower edge of the ED $(U,X)$ scans, so the window used in the main text, $U/t_1\ge 2$ and $X\ge 2$, lies within the valid region.

\section{Band folding by the $\sqrt{3}\times\sqrt{3}$ charge order and the $C=2$ sub-band}
\label{sec:sm-folding}

This section connects the Hall conductance $\sigma_{xy}=2\,e^2/h$ of the $\nu=1/3$ state to band folding by the $\sqrt{3}\times\sqrt{3}$ charge order at $\mathbf{Q}=K=\tfrac13\mathbf{b}_1+\tfrac23\mathbf{b}_2$.
Since $3\mathbf{Q}$ is a reciprocal lattice vector, the order triples the unit cell and folds the $C=+1$ lower Haldane band into three sub-bands, constrained only by the sum rule $C_1+C_2+C_3=+1$; the filled lowest sub-band is therefore not bounded by the parent band's $\lvert C\rvert=1$.
Here we construct a mean-field model that realizes such a Chern-number assignment, in the spirit of the folded-band description of quantum anomalous Hall crystals in Ref.~\cite{Sheng2024:Quantum},
\begin{equation}
    \hat{H}_{\text{MF}} = - t_1 \!\!\sum_{\langle i,j \rangle}\!\! \hat{c}^{\dagger}_{i} \hat{c}_{j}
    - t_2 \!\!\sum_{\langle\langle i,j\rangle\rangle}\!\! e^{i \nu_{ij}\phi}\, \hat{c}^{\dagger}_{i} \hat{c}_{j}
    + \hat{V} + \hat{W},
    \label{eq:sm-fold-model}
\end{equation}
the bare Haldane model of Eq.~(1) of the main text, with the $t_3$ term dropped as in the spin-polarized limit and the same band-topology parameters, $t_2/t_1=(\sqrt{43/3})/12$ and $\phi=\pi+\arccos(3\sqrt{3/43})$, together with the on-site potential $\hat{V}$ and the bond field $\hat{W}$ that the charge order generates, derived below.
Lengths are measured in the nearest-neighbor bond length, and we use the basis of the main text, $\mathbf{a}_1 = (\sqrt{3},\,0)$, $\mathbf{a}_2 = (\sqrt{3}/2,\,3/2)$ with the $B$ site at $\boldsymbol{\delta}_B = (\sqrt{3}/2,\,1/2)$, together with its dual $\mathbf{a}_i\cdot\mathbf{b}_j = 2\pi\delta_{ij}$.

\begin{figure}[tbp]
    \centering
    \includegraphics[width=.6\columnwidth]{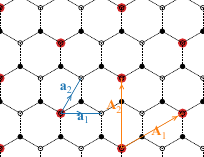}
    \caption{The $\sqrt{3}\times\sqrt{3}$ charge order at $\mathbf{Q}=K$. Red discs: the occupied $A$ sites, one per tripled cell, on a triangular lattice of spacing $\sqrt{3}\,\lvert\mathbf{a}_1\rvert$. Open/filled circles: the $A$/$B$ sublattices. Arrows: the primitive vectors $\mathbf{a}_{1,2}$ (blue) and the superlattice primitive vectors $\mathbf{A}_1 = \mathbf{a}_1 + \mathbf{a}_2$ and $\mathbf{A}_2 = 2\mathbf{a}_2 - \mathbf{a}_1$ (orange), the latter connecting occupied sites.}
    \label{fig:sm-cdw-pattern}
\end{figure}

{\it The ordered pattern.---}
The charge occupies one $\sqrt{3}\times\sqrt{3}$ subset of the $A$ sites (Fig.~\ref{fig:sm-cdw-pattern}), a triangular superlattice with primitive vectors $\mathbf{A}_1 = \mathbf{a}_1 + \mathbf{a}_2$ and $\mathbf{A}_2 = 2\mathbf{a}_2 - \mathbf{a}_1$. Its spacing lies outside the range of both the NN and 2NN repulsions, so the pattern costs no interaction energy, and its three translates correspond to the three ground states of the main text.
By inversion symmetry the same pattern on the $B$ sites is degenerate, and the finite-size ground states superpose the two; we model the $A$ choice, since both partners carry the same sub-band Chern numbers.

{\it On-site potential.---}
The tripled cell splits the $A$ sites into three classes, one per translate of the pattern. Every $B$ site has three $A$ neighbors, one from each class, and its six 2NN neighbors are all $B$ sites, which carry no modulation, so every $B$ site has the same potential.
The order therefore shifts the $B$ sublattice uniformly, and modulates only the $A$ sublattice: an $A$ site has six 2NN neighbors, three in each of the other two classes, so the 2NN density-density interaction of Eq.~(6) of the main text lowers the occupied class relative to them:
\begin{equation}
    \hat{V} = 2\Delta_A \sum_{i \in A} \cos(\mathbf{Q}\cdot\mathbf{r}_i)\, \hat{n}_{i},
    \label{eq:sm-fold-onsite}
\end{equation}
with $\mathbf{r}_i$ the site position and the modulation amplitude $\Delta_A < 0$.

{\it Bond field.---}
The order also modulates the 2NN hoppings.
Every 2NN bond has exactly one bridging site, i.e., the unique common neighbor of the pair. 
It lies on the opposite sublattice: an $A$ site for a $B$-$B$ bond, a $B$ site for an $A$-$A$ bond.
Only the $B$-$B$ bonds are therefore bridged by the modulated sublattice, and their modulation follows the occupation of that bridge:
\begin{align}
    \hat{W} &= -\sum_{\langle\langle i,j\rangle\rangle \in B}{\vphantom{\sum}}' \, w(\mathbf{r}_{m(ij)})\, e^{i\nu_{ij}\phi}\,
    \hat{c}^{\dagger}_{i} \hat{c}_{j} + \text{H.c.},
    \nonumber\\
    w(\mathbf{r}) &= 2\,\delta t_2 \cos\!\left(\mathbf{Q}\cdot\mathbf{r}\right),
    \label{eq:sm-fold-bond}
\end{align}
where $m(ij)$ is the link's bridge and the primed sum runs once over the $B$-sublattice 2NN links, oriented along their arrows in Fig.~1(a) of the main text. The link vector $\mathbf{v} \equiv \mathbf{r}_i - \mathbf{r}_j$ then takes the three values $\{-\mathbf{a}_1,\ \mathbf{a}_2,\ \mathbf{a}_1{-}\mathbf{a}_2\}$, a $C_3$ orbit, so the chirality $\nu_{ij}$ of Eq.~(1) of the main text is uniform, $e^{i\nu_{ij}\phi}=e^{i\phi}$.
The offsets $\boldsymbol{\rho}_{\mathbf{v}} \equiv \mathbf{r}_{m}-\mathbf{r}_{j}$ are set by geometry and the orientation above, with $\mathbf{Q}\cdot\boldsymbol{\rho}_{\mathbf{v}} = 4\pi/3$, $2\pi/3$, $0$ for the three links, so $\delta t_2$ is the bond field's only parameter.

{\it Folded Hamiltonian.---}
In momentum space, with $\psi_{\mathbf{k}} = (c_{\mathbf{k}A},\, c_{\mathbf{k}B})^{\top}$, both fields scatter $\mathbf{k}\to\mathbf{k}+\mathbf{Q}$,
\begin{align}
    \hat{V} + \hat{W} &= \sum_{\mathbf{k}} \psi^{\dagger}_{\mathbf{k}+\mathbf{Q}}\, V(\mathbf{k})\, \psi_{\mathbf{k}} + \text{H.c.},
    \nonumber\\
    V(\mathbf{k}) &= \operatorname{diag}\bigl(\Delta_A,\; V_B(\mathbf{k})\bigr),
    \nonumber\\
    V_B(\mathbf{k}) &= -\delta t_2\, e^{i\mathbf{Q}\cdot\boldsymbol{\delta}_B}
    \sum_{\mathbf{v}} e^{i\mathbf{Q}\cdot\boldsymbol{\rho}_{\mathbf{v}}}
    \bigl[ e^{i\phi - i(\mathbf{k}+\mathbf{Q})\cdot\mathbf{v}}
    \nonumber\\
    &\qquad\qquad\qquad\quad
    + e^{-i\phi + i\mathbf{k}\cdot\mathbf{v}} \bigr],
    \label{eq:sm-fold-vertex}
\end{align}
where $\mathbf{v}$ again runs over the three links. Writing $\mathbf{r}_m = \mathbf{R} + \boldsymbol{\delta}_B + \boldsymbol{\rho}_{\mathbf{v}}$ splits the bridge phase into three: the cell part $e^{i\mathbf{Q}\cdot\mathbf{R}}$ supplies the $\mathbf{k}\to\mathbf{k}+\mathbf{Q}$ scattering, and the other two are the factors $e^{i\mathbf{Q}\cdot\boldsymbol{\delta}_B}$ and $e^{i\mathbf{Q}\cdot\boldsymbol{\rho}_{\mathbf{v}}}$ of $V_B(\mathbf{k})$.
The three momenta then close under the scattering, and in the basis $\Psi_{\mathbf{k}} = (\psi_{\mathbf{k}},\, \psi_{\mathbf{k}+\mathbf{Q}},\, \psi_{\mathbf{k}+2\mathbf{Q}})^{\top}$ the Hamiltonian in the folded Brillouin zone is
\begin{equation}
    H_{\text{fold}}(\mathbf{k}) =
    \begin{pmatrix}
        H_0(\mathbf{k}) & V(\mathbf{k})^{\dagger} & V(\mathbf{k}+2\mathbf{Q}) \\
        V(\mathbf{k}) & H_0(\mathbf{k}+\mathbf{Q}) & V(\mathbf{k}+\mathbf{Q})^{\dagger} \\
        V(\mathbf{k}+2\mathbf{Q})^{\dagger} & V(\mathbf{k}+\mathbf{Q}) & H_0(\mathbf{k}+2\mathbf{Q})
    \end{pmatrix},
    \label{eq:sm-fold-H}
\end{equation}
with $H_0$ the $2\times2$ Bloch matrix of the two hopping terms of Eq.~(\ref{eq:sm-fold-model}), at $t_1=1$, so energies here carry the same scale as the rescaled ED energies.

The folded Hamiltonian $H_{\text{fold}}$ is periodic only up to a cyclic permutation $P$ of its momentum blocks, $H_{\text{fold}}(\mathbf{k}+\mathbf{G}) = P\, H_{\text{fold}}(\mathbf{k})\, P^{\dagger}$, where $\mathbf{G}$ runs over the reciprocal superlattice spanned by the duals of $\mathbf{A}_{1,2}$, $\mathbf{B}_1 = (2\mathbf{b}_1+\mathbf{b}_2)/3$ and $\mathbf{B}_2 = (\mathbf{b}_2-\mathbf{b}_1)/3$, whose Brillouin zone has one third the area of the parent zone and contains $\mathbf{Q} = \mathbf{B}_1+\mathbf{B}_2$.
We can also label the basis by position in the supercell: with $\mathbf{R}_c = c\,\mathbf{a}_1$ ($c=0,1,2$) the three cells, $\mathbf{k}_a = \mathbf{k}+a\mathbf{Q}$ and $\zeta = e^{2\pi i/3}$, the unitary $U_{ac}(\mathbf{k}) = e^{-i\mathbf{k}_a\cdot\mathbf{R}_c}/\sqrt{3}$, acting as the identity on the sublattice index, gives $\tilde{H} = U^{\dagger} H_{\text{fold}} U$ with $2\times2$ blocks
\begin{equation}
    \tilde{H}_{cc'} = \frac{e^{i\mathbf{k}\cdot(\mathbf{R}_c - \mathbf{R}_{c'})}}{3}
    \sum_{a} \zeta^{a(c-c')} \bigl[ H_0 + \zeta^{c} V + \zeta^{-c'} V^{\dagger} \bigr](\mathbf{k}_a),
    \label{eq:sm-fold-Hsc}
\end{equation}
now strictly periodic, $\tilde{H}(\mathbf{k}+\mathbf{G}) = \tilde{H}(\mathbf{k})$; the Chern numbers below are computed from it.

\begin{figure}[htbp]
    \centering
    \includegraphics{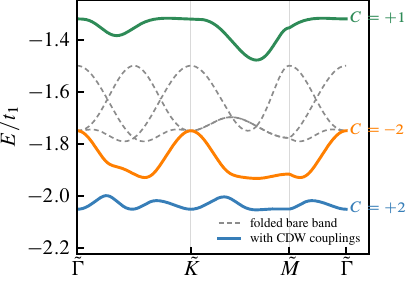}
    \caption{Sub-bands of the $\sqrt{3}\times\sqrt{3}$-folded lower Haldane band along $\tilde{\Gamma}$--$\tilde{K}$--$\tilde{M}$--$\tilde{\Gamma}$ of the folded Brillouin zone, with the charge-density-wave (CDW) couplings, $(\Delta_A,\, \delta t_2) = (-0.40,\, 0.15)$. 
    The three folded copies of the bare $C=+1$ band
    at $\Delta_A = \delta t_2 = 0$ are shown as dashed lines for comparison.
    }
    \label{fig:sm-folded-bands}
\end{figure}

{\it Sub-bands.---}
Figure~\ref{fig:sm-folded-bands} shows the three lowest sub-bands with and without the CDW couplings.
At $(\Delta_A,\, \delta t_2) = (-0.40,\, 0.15)$ they carry $(C_1, C_2, C_3) = (+2, -2, +1)$, and the lowest is separated from the rest by an indirect gap of $0.066\,t_1$, comparable to the many-body gap measured in the ED. At $\nu=1/3$ the model is therefore a Chern insulator with $\sigma_{xy} = 2\,e^2/h$.

\makeatletter%
\@ifundefined{auto@bib@empty}{}{\auto@bib@empty}%
\makeatother%